\documentclass[twocolumn]{aastex61}

\usepackage{xspace}
\usepackage{graphicx}
\usepackage{natbib}
\usepackage{amsmath}

\newcommand{\nraoblurb}{The National Radio Astronomy Observatory is
a facility of the National Science Foundation operated under cooperative
agreement by Associated Universities, Inc.}
\newcommand{\hide}[1]{}
\newcommand{\nexpo}[2]{\ensuremath{{#1}\times10^{#2}}\xspace}

\newcommand{\gl}{\ensuremath{\ell}\xspace}
\newcommand{\gb}{\ensuremath{{\it b}}\xspace}

\newcommand{\lb}{\ensuremath{(\gl,\gb)}\xspace}

\newcommand{\lbv}{\ensuremath{(\gl,\gb,v)}\xspace}

\newcommand{\kms}{\ensuremath{\,{\rm km\,s^{-1}}}\xspace}

\newcommand{\cmm}{\ensuremath{\,{\rm cm^{-3}}}\xspace}

\newcommand{\K}{\ensuremath{\,{\rm K}}\xspace}

\newcommand{\degree}{\ensuremath{^\circ}\xspace}

\newcommand{\arcmper}{\ensuremath{\rlap.{^{\prime}}}}

\newcommand{\hi}{{\rm H\,{\footnotesize I}}\xspace}
\newcommand{\hii}{{\rm H\,{\footnotesize II}}\xspace}
\newcommand{\cii}{{\rm C\,{\footnotesize II}}}
\newcommand{\nii}{{\rm N\,{\footnotesize II}}}

\newcommand{\hiea}{{\rm H\,{\footnotesize I}\,E/A}\xspace}

\shorttitle{Diffuse Ionized Gas in the Milky Way Disk}
\shortauthors{Luisi et al.}
\begin{document}

\title{Diffuse Ionized Gas in the Milky Way Disk}

\author[0000-0001-8061-216X]{Matteo Luisi}
\affiliation{Department of Physics and Astronomy, West Virginia University, Morgantown WV 26506, USA}
\affiliation{Center for Gravitational Waves and Cosmology, West Virginia University, Chestnut Ridge Research Building, Morgantown WV 26505, USA}

\author[0000-0001-8800-1793]{L. D. Anderson}
\affiliation{Department of Physics and Astronomy, West Virginia University, Morgantown WV 26506, USA}
\affiliation{Center for Gravitational Waves and Cosmology, West Virginia University, Chestnut Ridge Research Building, Morgantown WV 26505, USA}
\affiliation{Adjunct Astronomer at the Green Bank Observatory, P.O. Box 2, Green Bank WV 24944, USA}

\author[0000-0002-2465-7803]{Dana S. Balser}
\affiliation{National Radio Astronomy Observatory, 520 Edgemont Road, Charlottesville VA 22903-2475, USA}

\author[0000-0003-0640-7787]{Trey V. Wenger}
\affiliation{National Radio Astronomy Observatory, 520 Edgemont Road, Charlottesville VA 22903-2475, USA}
\affiliation{Astronomy Department, University of Virginia, P.O.~Box 400325, Charlottesville, VA 22904-4325, USA}

\author{T. M. Bania}
\affiliation{Institute for Astrophysical Research, Department of Astronomy, Boston University, 725 Commonwealth Ave., Boston MA 02215, USA}

\begin{abstract}
We analyze the diffuse ionized gas (DIG) in the first Galactic quadrant from $\ell=18\degree$ to $40\degree$ using radio recombination line (RRL) data from the Green Bank Telescope. These data allow us to distinguish DIG emission from \hii\ region emission and thus study the diffuse gas essentially unaffected by confusion from discrete sources. We find that the DIG has two dominant velocity components, one centered around $100\,\kms$ associated with the luminous \hii\ region W43, and the other centered around $45\,\kms$ not associated with any large \hii\ region. Our analysis suggests that the two velocity components near W43 may be caused by non-circular streaming motions originating near the end of the Galactic bar. At lower Galactic longitudes, the two velocities may instead arise from gas at two distinct distances from the Sun, with the most likely distances being $\sim$6\,kpc for the $100\,\kms$ component and $\sim$12\,kpc for the $45\,\kms$ component. We show that the intensity of diffuse {\it Spitzer} GLIMPSE 8.0\,\micron\ emission caused by excitation of polyaromatic hydrocarbons (PAHs) is correlated with both the locations of discrete \hii\ regions and the intensity of the RRL emission from the DIG. This implies that the soft ultra-violet photons responsible for creating the infrared emission have a similar origin as the harder ultra-violet photons required for the RRL emission. The 8.0\,\micron\ emission increases with RRL intensity but flattens out for directions with the most intense RRL emission, suggesting that PAHs are partially destroyed by the energetic radiation field at these locations.
\end{abstract}


\section{Introduction}
First proposed by \citet{hoyle63}, the warm interstellar medium (WIM) is a widespread component of the interstellar medium (ISM) with density $\sim$0.1\cmm\ and temperatures from 6000 to 10000\,K \citep[see][and references therein]{haffner09}. At the upper end of this temperature range, the WIM is nearly fully ionized, with a hydrogen ionization ratio $n(\textnormal{H}^+)/n(\textnormal{H}^0) \geq 13$ \citep{reynolds98}. Thus, the WIM is also known as the ``Diffuse Ionized Gas" (DIG). Despite its low density, $\sim$80--90\% of the total free-free emission in our Galaxy is thought to come from the DIG.

Though the exact mechanisms are still unknown, it is believed that the DIG maintains its ionization from O-type stars, whose UV radiation leaks out of the \hii\ regions surrounding them and into the ISM \citep{reynolds84, ferguson96, zurita02}. \citet{Murray2010} confirmed that a large number of ionizing photons are leaking from \hii\ regions. \citet{Anderson2015} derived an ionizing radiation leaking fraction of $\sim$25\% for the bubble \hii\ region RCW\,120 using H$\alpha$ data at 656\,nm. They also showed that the photodissociation region (PDR) surrounding the \hii\ region has distinct ``holes" through which photons can escape into the ISM. This suggests that PDRs are generally not homogeneous. Recently, we showed that the non-uniform PDR surrounding the compact \hii\ region NGC\,7538 allows radiation to escape preferentially along a single direction \citep{Luisi2016}. We calculated a leaking fraction $f_R = 15 \pm 5$\% of the radio continuum emission. This leaking emission appears spatially confined within an additional, more distant PDR boundary around NGC\,7538 and thus seems to only affect the local ambient medium. Results suggest, however, that giant \hii\ regions such as W43 may have a much larger effect in maintaining the ionization of the DIG and despite their small numbers may be the dominant source of ionizing radiation in the ISM \citep[see][]{Zurita2000}.

Together with \hii\ regions and PDRs, the DIG is a major source of radio recombination line (RRL) emission. Consequently, RRL observations have been used to map its spatial and velocity distribution. Compared to studies of optical emission lines, specifically H$\alpha$ \citep[e.g., the WHAM survey,][]{Haffner1999}, RRL observations have the advantage of essentially being free from extinction due to interstellar dust. Their disadvantage is reduced sensitivity, restricting RRL detections to gas with higher emission measure than that traced by H$\alpha$. The fully-sampled 1.4\,GHz RRL survey of \citet{alves10, alves12, alves15} mapped the plane of the Galaxy at a spatial resolution of $14\arcmper4$.  They were, however, unable to distinguish the contributions from discrete \hii\ regions and the DIG for most sight lines. The observing method of the fully-sampled SIGGMA RRL survey \citep{liu13} partially filters out the emission from the DIG. Finally, \citet{Roshi2001} observed the Galactic plane in RRLs near 327\,MHz from $-28\degree < \ell < 89\degree$. Despite the low resolution of $\sim$$2\degree$, they obtain an upper limit of 12,000\,K for the electron temperature of the gas and suggest that the emission originates from low-density ionized gas forming \hii\ region envelopes.

With the emergence of high-sensitivity RRL surveys, the DIG has been serendipitously detected in observations of discrete \hii\ regions \citep[see][and references therein]{Anderson2015a}. In the Green Bank Telescope \hii\ Region Discovery Survey \citep[GBT HRDS, see][]{Bania2010, Anderson2011} we identified multiple RRL velocity components in $\sim$30\% of all observed targets. This fraction is too large to be caused by multiple discrete \hii\ regions along the line of sight \citep[see][]{Anderson2015a}. We thus infer that the RRL emission at these locations is usually composed of emission from a discrete source and emission from the DIG \citep{Anderson2015a}.

Here, we use data from past observations \linebreak
\citep{Anderson2011, Anderson2015a} and previously unpublished data for directions either known to be devoid of discrete \hii\ regions, or in directions where the \hii\ region emission can be distinguished from that of the DIG (see \S 2 for details on how we distinguish between these two components). This gives us an irregularly-spaced grid of pointings, for which we can extract the intensity and velocity of only the DIG. The advantage of our strategy is that the beam size is relatively small ($82\arcsec$) compared with typical spacings between discrete \hii\ regions so the emission at each pointing is not contaminated with \hii\ region emission.  The disadvantage of course is that the \lb-space is not fully sampled. By distinguishing the emission from discrete \hii\ regions and the DIG, our data allows us to essentially filter out \hii\ region emission entirely and map only the diffuse component. This gives us an advantage over previous RRL surveys \citep{liu13, alves15} as these are at least partially contaminated by \hii\ region emission. With this analysis we are able to investigate the relationship between discrete \hii\ regions and the diffuse gas, and test our hypothesis that large \hii\ regions are dominant in maintaining the ionization of the DIG \citep{Zurita2000, Luisi2016}.

\section{DIG RRL Emission}
Our RRL emission data were taken with the Auto-Correlation Spectrometer (ACS) on the National Radio Astronomy Observatory Green Bank Telescope (GBT).  We observed a total of 254 directions between $\ell=18\degree$ and $40\degree$ and $|b| < 1\degree$ which yielded 379 sets of line parameters for the DIG. Our data come from two previously published sources, \citet{Anderson2011} and \citet{Anderson2015a}, and one previously unpublished source (see below). \citet{Anderson2011} contains directions coincident with \hii\ regions, as defined by 8\,$\mu$m \emph{Spitzer} GLIMPSE emission, for which the diffuse gas velocity can be distinguished from the \hii\ region velocity (these data include 98 pointings with 116 sets of diffuse line parameters). The process of distinguishing the diffuse gas velocities from \hii\ region velocities is described in \citet{Anderson2015a}. We use previous GBT observations, analyze the derived electron temperature for each velocity component, and search for the molecular emission or carbon recombination lines associated with one RRL component. Sight lines that that do not pass within the 8\,$\mu$m-defined \hii\ region PDR are always considered ``diffuse." \citet{Anderson2015a} also includes such directions devoid of discrete \hii\ regions which allows us to directly sample the DIG without confusion (135 pointings; 237 sets of line parameters). 

We also incorporate observations taken near the giant \hii\ region W43 (21 pointings; 26 sets of diffuse line parameters) which we have not published previously. Here, we use our HRDS data to distinguish the diffuse gas velocity from the \hii\ region velocity. If the observed direction is spatially coincident with a known \hii\ region, we assume that the velocity component closest to the \hii\ region velocity is due to the \hii\ region itself. We summarize these data in Table~\ref{tab:w43}, which lists the source, the Galactic longitude and latitude, the hydrogen line intensity, the FWHM line width, the local standard of rest (LSR) velocity, and the rms noise in the spectrum, including all corresponding 1$\sigma$ uncertainties of the Gaussian fits. For directions with multiple velocity components detected along the line of sight, the source names are given additional letters, ``a," ``b," or ``c," in order of decreasing peak line intensity. Velocity components that are due to discrete \hii\ regions are marked with an asterisk in the table and are not used for our data analysis. For each observed direction, we simultaneously measured 7 Hn$\alpha$ RRL transitions in the 9\,GHz band, H87$\alpha$ to H93$\alpha$, using our standard techniques \citep{Bania2010, Balser2011, Anderson2011}, and averaged all spectra together to increase the signal-to-noise ratio using TMBIDL\footnote{V7.1, see https://github.com/tvwenger/tmbidl.git.} \citep{Bania2014}. We assume that the brightest line emission from the DIG is due to hydrogen and fit a Gaussian model to each line profile. We use the line intensities, full width at half maximum (FWHM) values, and LSR velocities derived from the Gaussian fits for all further analysis.

\begin{deluxetable*}{lcccccccccc}
\tablewidth{0pt}
\tablecaption{RRL Emission Near W43 \label{tab:w43}}
\tablehead{Source  & \colhead{$\ell$} & \colhead{$b$} & \colhead{$T_L$} & \colhead{$\sigma T_L$} & \colhead{$\Delta V$} & \colhead{$\sigma \Delta V$} & \colhead{$V_{\rm LSR}$} & \colhead{$\sigma V_{\rm LSR}$} & \colhead{rms} & \colhead{Note\tablenotemark{a}}\\
  & \colhead{(degree)} & \colhead{(degree)} & \colhead{(mK)} & \colhead{(mK)} & \colhead{(\kms)} & \colhead{(\kms)} & \colhead{(\kms)} & \colhead{(\kms)} & \colhead{(mK)} & \colhead{}}
\startdata
G030.400+0.180 &   30.400 & +0.180 & $\phn \phn \phn 8.2$  &  $0.1$  &  $58.7$  &  $1.5$  &  \phn 88.1  &  $0.5$ & 0.7 & \\
G030.570$-$0.230 &   30.570 & $-$0.230 & $\phn \phn 38.8$  &  $0.3$  &  $19.3$  &  $0.2$  &  \phn 88.1 &  $0.1$ & 0.4 & * \\
G030.570+0.090a &   30.570 & +0.090 & $\phn \phn 27.6$  &  $0.3$  &  $23.2$  &  $0.3$  &  \phn 42.4  &  $0.1$ & 1.7 & \\
G030.570+0.090b &   30.570 & +0.090 & $\phn \phn 22.0$  &  $0.3$  &  $24.9$  &  $0.4$  &  104.9  &  $0.2$ & 1.7 & \\
G030.650$-$0.150a &   30.650 & $-$0.150 & $\phn \phn 53.8$  &  $0.2$  &  $22.0$  &  $0.1$  &  \phn 96.0  &  $0.1$ & 0.5 & \\
G030.650$-$0.150b &   30.650 & $-$0.150 & $\phn \phn 24.1$  &  $0.2$  &  $18.4$  &  $0.2$  &  119.3  &  $0.1$ & 0.5 & \\
G030.740$-$0.060 &   30.740 & $-$0.060 & $\phn 617.6$  &  $1.5$  &  $24.2$  &  $0.1$  &  \phn 90.7 &  $0.1$ & 2.1 & * \\
G030.740+0.010 &   30.740 & +0.010 & $\phn 612.5$  &  $0.7$  &  $29.5$  &  $0.1$  &  \phn 91.1 &  $0.1$ & 2.3 & * \\
G030.740+0.100a &   30.740 & +0.100 & $\phn \phn 34.8$  &  $0.3$  &  $21.5$  &  $0.2$  &  120.9  &  $0.1$ & 1.6 & \\
G030.740+0.100b &   30.740 & +0.100 & $\phn \phn 24.3$  &  $0.3$  &  $19.8$  &  $0.3$  & \phn 88.0  &  $0.1$ & 1.6 & * \\
G030.740+0.100c &   30.740 & +0.100 & $\phn \phn \phn 9.9$  &  $0.3$  &  $20.5$  &  $0.7$  &  \phn 39.9  &  $0.3$ & 1.6 & \\
G030.740+0.180a &   30.740 & +0.180 & $\phn \phn 12.8$  &  $0.2$  &  $46.7$  &  $1.1$  &  \phn 96.6  &  $0.4$ & 1.5 & * \\
G030.740+0.180b &   30.740 & +0.180 & $\phn \phn 12.5$  &  $0.3$  &  $21.6$  &  $0.7$  &  \phn 39.5  &  $0.3$ & 1.5 & \\
G030.740+0.260a &   30.740 & +0.260 & $\phn \phn 50.3$  &  $0.4$  &  $20.7$  &  $0.2$  &  100.5  &  $0.1$ & 1.3 & * \\
G030.740+0.260b &   30.740 & +0.260 & $\phn \phn 13.1$  &  $0.4$  &  $19.7$  &  $0.7$  &  \phn 37.0  &  $0.3$ & 1.3 & \\
G030.740+0.280a &   30.740 & +0.280 & $\phn \phn 64.2$  &  $0.4$  &  $19.7$  &  $0.1$  &  100.8  &  $0.1$ & 1.2 & * \\
G030.740+0.280b &   30.740 & +0.280 & $\phn \phn \phn 9.9$  &  $0.4$  &  $20.3$  &  $0.8$  &  \phn 37.2  &  $0.4$ & 1.2 & \\
G030.740+0.300a &   30.740 & +0.300 & $\phn \phn 34.0$  &  $0.3$  &  $16.9$  &  $0.2$  &  102.4  &  $0.1$ & 1.3 & * \\
G030.740+0.300b &   30.740 & +0.300 & $\phn \phn 11.3$  &  $0.2$  &  $23.5$  &  $1.1$  &  \phn 79.0  &  $0.5$ & 1.3 & \\
G030.740+0.300c &   30.740 & +0.300 & $\phn \phn \phn 7.6$  &  $0.2$  &  $22.3$  &  $0.8$  &  \phn 36.7  &  $0.3$ & 1.3 & \\
G030.740+0.350a &   30.740 & +0.350 & $\phn \phn 15.8$  &  $0.3$  &  $14.7$  &  $0.5$  &  \phn 78.2  &  $0.2$ & 1.2 & \\
G030.740+0.350b &   30.740 & +0.350 & $\phn \phn 15.7$  &  $0.3$  &  $21.6$  &  $0.6$  &  102.3  &  $0.2$ & 1.2 & \\
G030.740+0.350c &   30.740 & +0.350 & $\phn \phn \phn 3.6$  &  $0.2$  &  $35.4$  &  $3.7$  &  \phn 39.5  &  $1.2$ & 1.2 & \\
G030.740+0.430 &   30.740 & +0.430 & $\phn \phn \phn 8.8$  &  $0.3$  &  $28.0$  &  $1.0$  &  \phn 98.8  &  $0.4$ & 1.7 & \\
G030.740+0.510a &   30.740 & +0.510 & $\phn \phn \phn 8.2$  &  $0.2$  &  $34.8$  &  $0.8$  &  \phn 92.1  &  $0.3$ & 1.2 & \\
G030.740+0.510b &   30.740 & +0.510 & $\phn \phn \phn 5.7$  &  $0.3$  &  $10.7$  &  $0.6$  &  \phn 18.9  &  $0.3$ & 1.2 & \\
G030.780$-$0.020 &   30.780 & $-$0.020 & $2179.2$  &  $2.6$  &  $31.4$  &  $0.1$  &  \phn 91.7 &  $0.1$ & 4.8 & * \\
G030.780+0.010 &   30.780 & +0.010 & $\phn 227.1$  &  $0.4$  &  $31.4$  &  $0.1$  &  \phn 92.6 &  $0.1$ & 2.2 & * \\
G030.820$-$0.060 &   30.820 & $-$0.060 & $\phn 280.0$  &  $0.4$  &  $29.0$  &  $0.1$  &  106.3 &  $0.1$ & 1.8 & * \\
G030.820+0.180a &   30.820 & +0.180 & $\phn \phn 14.0$  &  $0.2$  &  $18.5$  &  $0.3$  &  \phn 36.4  &  $0.1$ & 0.8 & \\
G030.820+0.180b &   30.820 & +0.180 & $\phn \phn 11.3$  &  $0.1$  &  $38.7$  &  $0.5$  &  \phn 99.6  &  $0.2$ & 0.8 & * \\
G030.900$-$0.060a &   30.900 & $-$0.060 & $\phn \phn 33.1$  &  $1.7$  &  $18.3$  &  $0.4$  &  107.3  &  $0.4$ & 1.2 & \\
G030.900$-$0.060b &   30.900 & $-$0.060 & $\phn \phn 14.5$  &  $0.8$  &  $23.5$  &  $1.9$  &  \phn 90.0  &  $1.3$ & 1.2 & \\
G030.900$-$0.060c &   30.900 & $-$0.060 & $\phn \phn \phn 4.6$  &  $0.2$  &  $31.2$  &  $1.8$  &  \phn 45.7  &  $0.7$ & 1.2 & \\
G030.900+0.340a &   30.900 & +0.340 & $\phn \phn \phn 7.3$  &  $0.1$  &  $37.3$  &  $0.7$  &  103.4  &  $0.3$ & 0.8 & \\
G030.900+0.340b &   30.900 & +0.340 & $\phn \phn \phn 2.8$  &  $0.1$  &  $28.6$  &  $1.7$  &  \phn 38.4  &  $0.7$ & 0.8 & \\
G031.070$-$0.150a &   31.070 & $-$0.150 & $\phn \phn 13.8$  &  $0.2$  &  $28.6$  &  $0.4$  &  \phn 99.0  &  $0.2$ & 0.9 & \\
G031.070$-$0.150b &   31.070 & $-$0.150 & $\phn \phn \phn 4.3$  &  $0.2$  &  $25.8$  &  $1.4$  &  \phn 26.2  &  $0.6$ & 0.9 & \\
\enddata
\tablenotetext{$a$}{\,RRL components associated with discrete \hii\ regions are marked with an asterisk (*, see text).}
\end{deluxetable*}

\begin{figure}[t]
  \centering
  \includegraphics[width=0.5\textwidth]{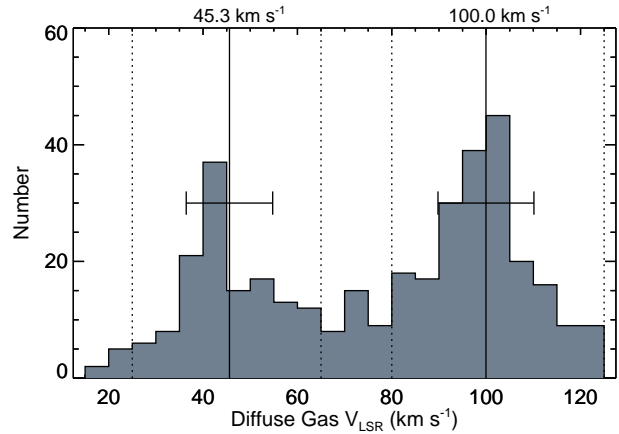}
  \caption{Velocity distribution of the DIG. The LSR velocities are derived from Gaussian fits to our RRL data. The dotted lines are defined by eye and show the velocity ranges that contain the majority of the diffuse gas velocities: the $45$\,\kms\ component at $25-65$\,\kms\ and the $100$\,\kms\ component at $80-125$\,\kms. The solid lines and error bars indicate the mean velocities within these ranges and their standard deviations.\label{fig:vlsr}}
\end{figure}

\section{Discussion}
\subsection{The Galactic Location of the DIG}
Over the longitude range considered here, the DIG emission is concentrated near two velocities, $45$\,\kms\ and $100$\,\kms\ (Figure~\ref{fig:vlsr}). This suggests that within our observed Galactic longitude range the DIG itself is located at two distinct distances, assuming that the diffuse gas in each velocity range can be assigned a single distance. We summarize the DIG emission properties in Table~\ref{tab:par}.

Just as for the discrete sources, however, this diffuse gas also suffers from the kinematic distance ambiguity (KDA). Unfortunately, we cannot use the \hi\ emission/absorption (\hiea) method \citep{kuchar94, kolpak03, Anderson2012a} for the DIG, both because it is faint and also because of the difficulty in finding a suitable ``off'' position.  Only massive stars can produce ionizing photons energetic enough to create and maintain the DIG \citep[e.g.,][]{reynolds84}.  We can therefore potentially determine the kinematic distance for the diffuse gas by associating it with massive star formation tracers that have their KDA resolved: massive \hii\ regions, molecular gas, and cold \hi. 

Below, we attempt to find the distance to the two observed velocity components of the DIG by resolving their KDA. In \S 3.1.1 and \S 3.1.2 we assume that each velocity component can be assigned a single distance from the Sun. In \S 3.1.3 we explore the possibility of the two observed velocity components being due to interacting gas clouds at the same distance from the Sun.

\subsubsection{The 45\,km\,s$^{-1}$ Gas Component}

The KDA leads to two possible distance ranges for each velocity range.  The $45$\,\kms\ gas could be at either $1.7 - 3.7$\,kpc\ or $10.6 - 12.7$\,kpc, if we assume $\lb = (30\degree, 0\degree$), and use the \citet{Reid2014} rotation curve (see Table~\ref{tab:par}).

\begin{deluxetable}{lcc}
\tabletypesize{\footnotesize}
\tablewidth{0pt}
\tablecaption{DIG Parameters \label{tab:par}}
\tablehead{Velocity range  & \colhead{$45$\,\kms} & \colhead{$100$\,\kms}}
\startdata
Number of RRL components (total) & 128 & 211\\
Number of RRL components (on)\tablenotemark{a} & 33 & 63\\
Number of RRL components (off) & 95 & 148\\
\hline 
Mean velocity (\kms) & 45.3 & 100.0\\
Median velocity (\kms) & 43.0 & 99.4\\
Std.~Dev.~velocity (\kms) & 9.2 & 10.1\\
\hline
Mean $T_{\rm A}$ (mK) & 12.5 & 16.8\\
Median $T_{\rm A}$ (mK) & 9.8 & 13.6\\
Std.~Dev.~$T_{\rm A}$ (mK) & 9.6 & 11.2\\
\hline
Near distance (kpc) & $1.7-3.7$ & $4.4-7.2$\\
Far distance (kpc) & $10.6-12.7$ & $7.2-10.0$\\
Assumed distance (kpc) & $\sim$12 & $\sim$6\\
\hline
Total integrated flux (Jy) & 172.8 & 246.0\\
Total integrated flux (Jy)\tablenotemark{b} & 118.3 & 220.4 \\
\enddata
\tablenotetext{$a$}{\,``on" and ``off" correspond to directions coincident with \hii\ \\
regions (on), and directions devoid of discrete \hii\ regions (off).}
\tablenotetext{$b$}{\,From \citet{alves15}.}
\end{deluxetable}

Assuming that the DIG is maintained by massive stars, we can use the ionization rate of \hii\ regions as a tracer to determine the distance to the DIG. In the range $\ell = 18\degree$ to $40\degree$, there are 205 \hii\ regions with velocities between 25 and 65\,\kms, and 127 of these have kinematic distance ambiguity resolutions \citep[KDARs;][The WISE Catalog of Galactic \hii\ Regions, Version 1.4\footnote{see http://astro.phys.wvu.edu/wise/}]{anderson14}. The total radio flux density of the 94 regions at the far kinematic distance is 10.84\,Jy, whereas the total flux density of the 33 regions at the near kinematic distance is only 1.09\,Jy. We estimate the ionization rate for each region using our HRDS data \citep{Rubin1968,Anderson2010a} by
\begin{equation} 
N_{\rm ly} \approx 4.76 \times 10^{48} \left( \frac{S_{\nu}}{\textnormal{Jy}} \right) \left( \frac{T_{\rm e}}{\textnormal{K}} \right) ^{-0.45} \left( \frac{\nu}{\textnormal{GHz}} \right) ^{0.1} \left( \frac{d}{\textnormal{kpc}} \right) ^2,
\end{equation} 
where $N_{\rm ly}$ is the ionization rate, the number of emitted Lyman Continuum ionizing photons per second, $S_{\nu}$ is the radio flux density of the \hii\ region, $T_{\rm e}$ is the electron temperature, $\nu = 1.4$\,GHz is the observed frequency \citep[see][]{Anderson2011}, and $d$ is the distance to the region. We assume a constant $T_{\rm e} = 10^4$\,K and sum the contribution for each individual region to find the total $N_{\rm ly}$ for \hii\ regions at the far and near kinematic distance. This estimate yields $N_{\rm ly} = \nexpo{1.06}{50} \, \textnormal{s} ^{-1}$ for the far distance and only $N_{\rm ly} = \nexpo{5.80}{47} \, \textnormal{s} ^{-1}$ for the near distance. This suggests that most of the DIG near $45$\,\kms\ is also at the far kinematic distance.

There is also over twice as much total CO gas at the far kinematic distance for clouds in the velocity range 25 to 65\,\kms compared to the near distance.  The average near GRS cloud CO luminosity from \citet{roman-duval09} in units of $10^4\K\,\kms\,{\rm pc}^{-2}$ is 0.23 with a standard deviation of 0.31, while it is 1.4 with a standard deviation of 1.7 for the far GRS clouds.  The total CO luminosity for the near clouds is $\nexpo{3.4}{5} \K\,\kms\,{\rm pc}^{-2}$, while it is $\nexpo{8.0}{5} \K\,\kms\,{\rm pc}^{-2}$ for the far GRS clouds. This again supports the $45$\,\kms\ DIG being at the far kinematic distance, if it is indeed associated with the molecular gas traced by CO emission that will continue to form massive stars.

\begin{figure}
  \centering
  \includegraphics[width=0.48\textwidth]{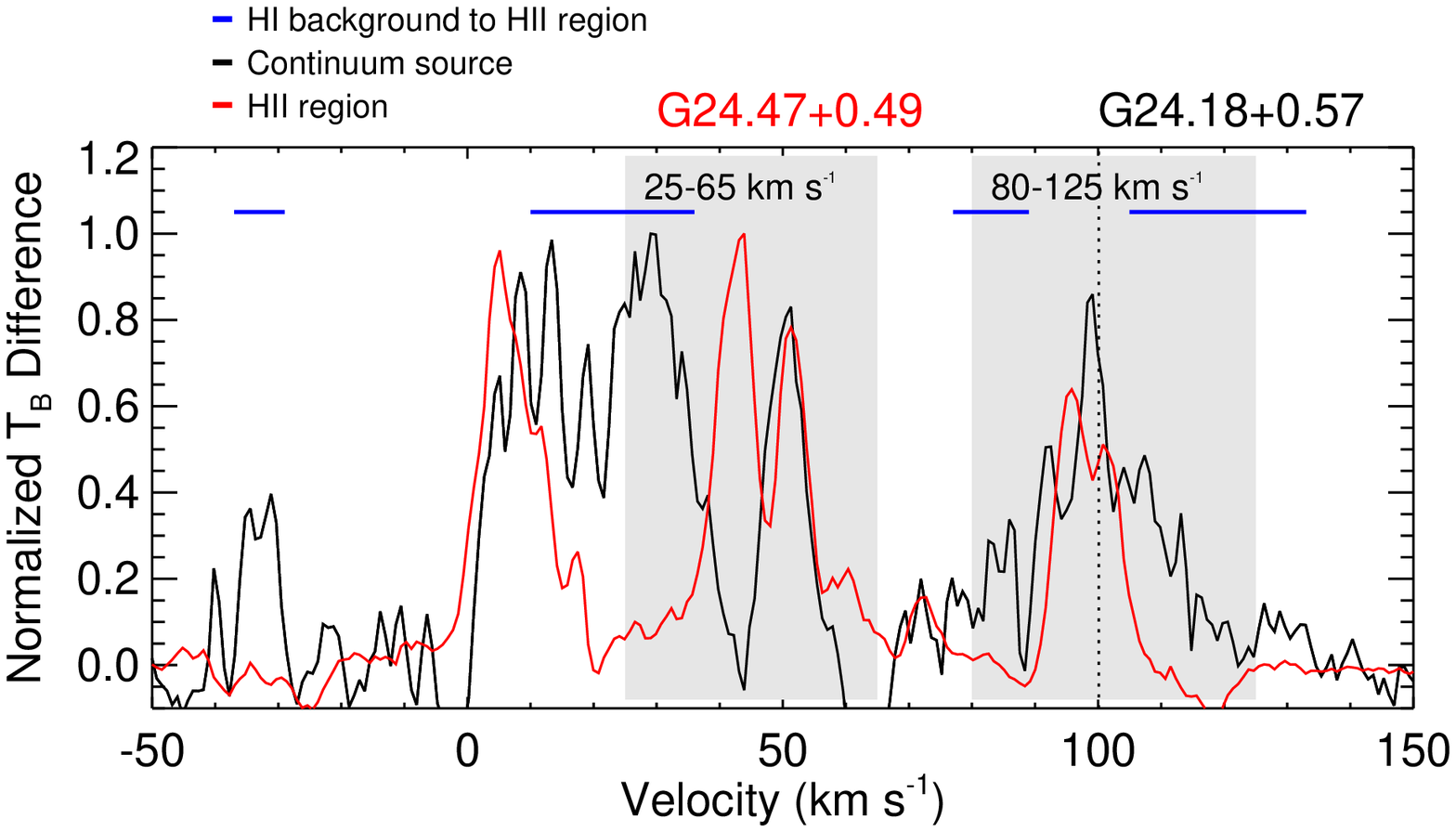}
  \includegraphics[width=0.48\textwidth]{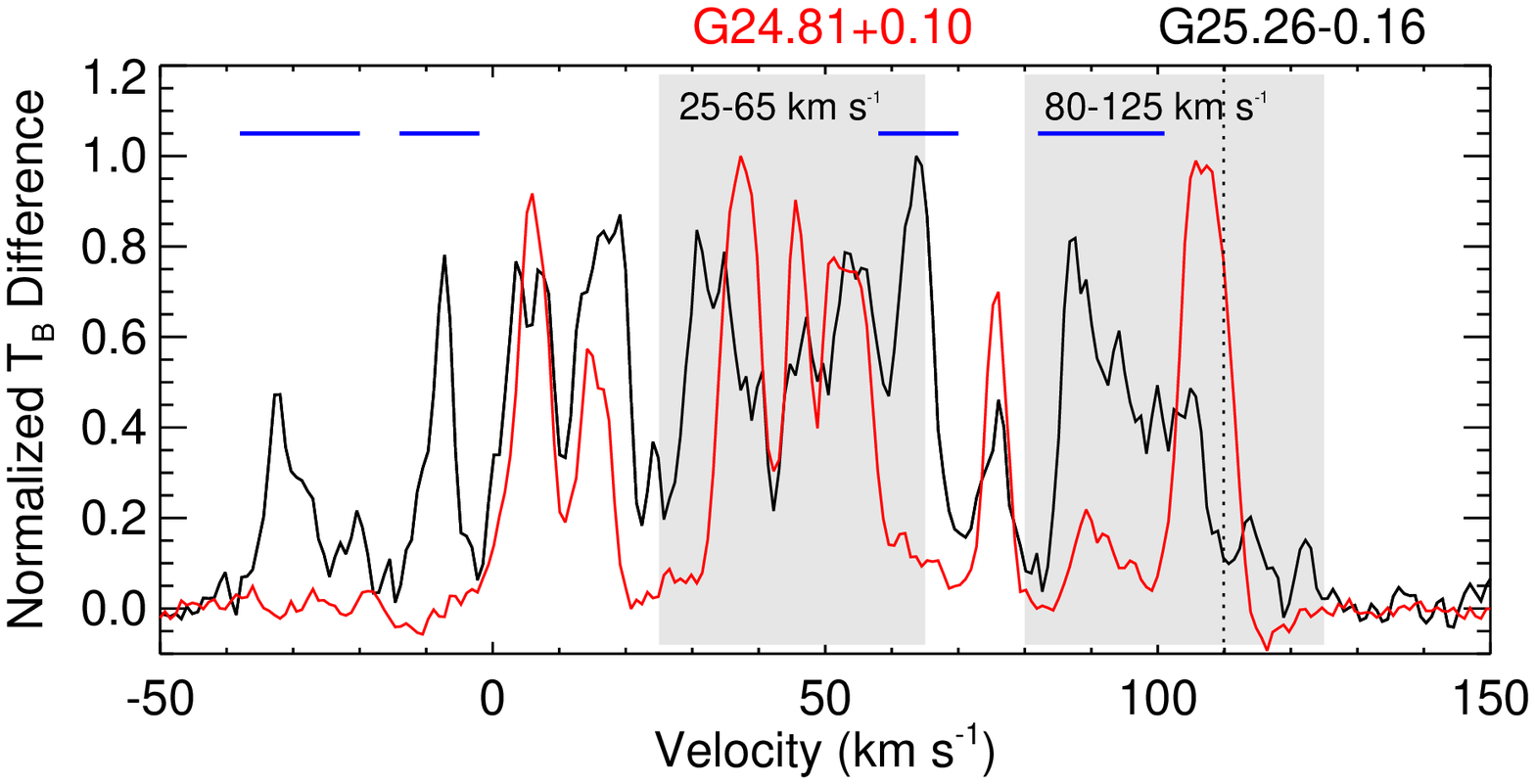}
  \includegraphics[width=0.48\textwidth]{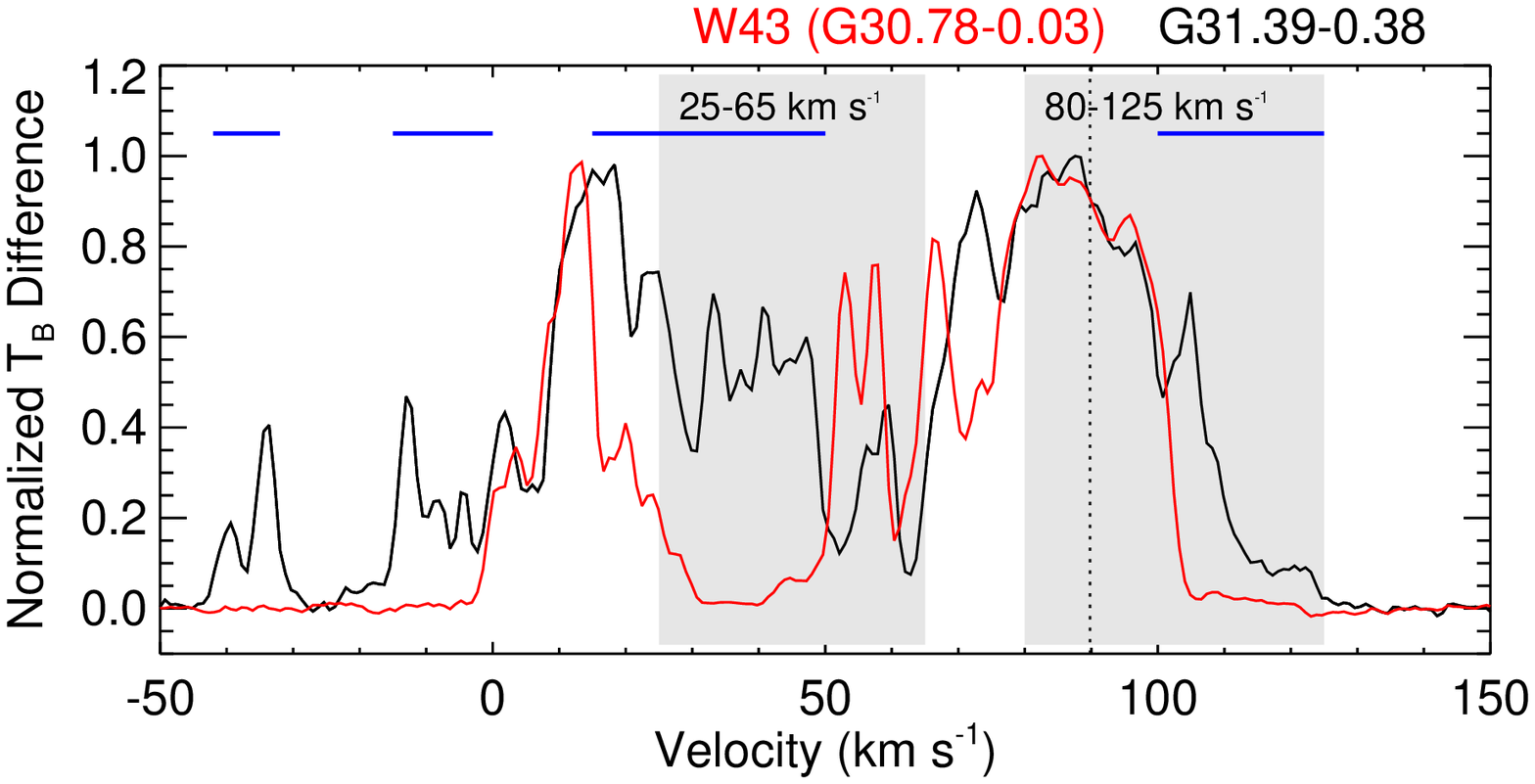}
  \caption{\hi\ absorption spectra towards \hii\ regions and extragalactic radio sources to investigate the KDAR of the DIG. The difference in VGPS main beam temperature, normalized by the maximum observed temperature, is shown between on- and off-target directions for \hii\ regions (red curves) and extragalactic radio continuum sources (black curves).  The top panel shows the \hii\ region G24.47$+$0.49 and the extragalactic source G24.18$+$0.57, the middle panel shows G24.81$+$0.10 and the extragalactic source G25.26$-$0.16, and the bottom panel shows W43 at $\lb \simeq (30.8\degree, 0.0\degree)$ and the extragalactic source G31.39$-$0.38. The \hii\ region RRL velocities are shown with dotted lines. Shaded regions indicate the two dominant velocity ranges of the DIG emission. Continuum source spectra showing absorption not present in the \hii\ region spectra should be background to the \hii\ region; these velocities are marked with horizontal blue lines at the top of each plot.  This analysis implies that most of the \hi\ below 50\,\kms\ near W43 is at the far kinematic distance. \label{fig:onoff}}
\end{figure}

Finally, we investigate the location of the cold \hi\ gas using the \hiea\ method. Only cold \hi\ foreground to a radio continuum source will cause \hi\ absorption, assuming that \hi\ self-absorption is negligible. The \hi\ spectrum toward an extragalactic radio continuum source can show absorption for all \hi\ along the line of sight, while for Galactic sources the \hi\ spectrum can only show absorption up to the source velocity. Comparing the \hi\ spectra toward nearby extragalactic and Galactic \hii\ region pairs can therefore tell us about the \hi\ distribution. If cold \hi\ gas is foreground to the \hii\ region, we expect to see absorption in both spectra. Cold \hi\ beyond the \hii\ region, however, will only show absorption in the spectrum toward the extragalactic source.

Here we use the Very Large Array Galactic Plane Survey \citep[VGPS; see][]{stil06} spectral line data to compare the \hiea\ spectrum for three \hii\ regions (G24.47+0.49, G24.81+0.10, and W43) with velocities near $100$\,\kms. All three have nearby (within $\sim$$40\arcmin$) extragalactic radio continuum sources. Figure~\ref{fig:onoff} shows the difference between on- and off-target directions for the \hii\ regions and extragalactic radio sources, where the on- and off-positions are separated by 6\arcmin. The \hii\ regions are located either foreground or background to the $45\,\kms$ gas, depending on their KDARs. As a result, \hi\ gas at velocities showing extragalactic absorption which is not present in the \hii\ region spectra should be background to the \hii\ region. This analysis implies that most of the \hi\ below 50\,\kms near W43 is at the far kinematic distance. The first \hi\ spectrum pair (G24.47+0.49) extracted near $\ell = 24$\degree shows partial absorption near $45\,\kms$ that is inconsistent with the absorption features seen in the second pair near $\ell = 24$\degree (G24.81+0.10). Therefore, we can not assign a single distance to the \hi\ near the $\ell = 24$\degree region. These results are somewhat ambiguous, however, since the separation between the line of sight towards the \hii\ regions and the extragalactic continuum sources are probing different \hi\ volumes. 

Since both the total \hii\ region ionization rate and the fraction of molecular gas are greater at the far distance, we favor the conclusion that most of the 45\,\kms\ diffuse gas is at its far kinematic distance of $\sim$12\,kpc as well. This is a simplified assumption and does not take into account the existence of additional gas at the other distance.

\subsubsection{The 100\,km\,s$^{-1}$ Gas Component}
The possible distance range for the $100$\,\kms\ gas is $4.4 - 10.0$\,kpc\ for $\lb\ = (30\degree, 0\degree$).  Because the molecular gas and massive star formation for the $80-100$\,\kms\ \lbv\ locus is associated with W43 \citep[e.g.,][]{nguyenlong11} at a distance of 5.49$^{+0.39}_{-0.34}$~kpc \citep{Zhang2014}, we assume throughout the remainder of this paper that the $100$\,\kms\ DIG is at a distance of $\sim$6\,kpc. Recently, \citet{Langer2017} observed the DIG along 18 lines of sight between $\ell = 30$\degree and $32$\degree using the [\cii] 158\,$\mu$m and [\nii] 205\,$\mu$m fine structure lines. They find a strong line component near $\sim$115\kms and argue that this component is due to DIG emission associated with the inner edge of the Scutum spiral arm tangency at a distance of $\sim$7\,kpc. Even if our assumption that the gas is at the distance of $\sim$6\,kpc is poor, our conclusions below are largely unaffected.

\vspace{16pt}\subsubsection{Interacting Gas Clouds?}
Our detection of the DIG in two separate velocity ranges suggests that each velocity range is primarily located at either its near or its far kinematic distance. If the two velocity components are indeed interacting, we would expect to observe an interaction signature between them. Such an interaction signature has been suggested by \citet{Beuther2012a} for the $^{13}$CO(2$-$1) emission near the W43 region, as well as for dense gas tracers like N$_2$H$^+$. In the Milky Way, however, this picture is further complicated by the vicinity of the $45$\,\kms component to the Galactic bar and the Scutum arm. Using an extragalactic counterpart to the W43 region, \citet{Beuther2017} argue that gas buildup near the bar/spiral arm interface, where W43 is located, is likely due to crossings between different orbit families. They posit that the observed velocities in the bar/spiral arm interface of NGC\,3627 are primarily due to interacting gas clouds.

If we assume that the observed velocities toward the $\ell \sim 30\degree$ region are due to interacting gas clouds at a single distance, we can use the method described by \citet{Beuther2017} to estimate the expected gas velocities observed along the line of sight and compare these with our observations. The simplest approximation assumes that the observed diffuse gas towards W43 is located at the tip of the Galactic bar, and that the two observed velocity signatures are due to the unperturbed, purely circular gas motion around the Galactic center and gas streaming motions along the bar, respectively. Using the \citet{Reid2014} rotation curve, we find a circular gas velocity, $V_C \sim 230$\,\kms for the observed diffuse gas towards W43. This corresponds to a velocity component along the line of sight of 91\,\kms, almost identical to the observed velocity of 89.8\,\kms for W43 itself. The perturbed velocity component due to streaming motions can be described by determining the bar perturbation to the gravitational potential \citep[see][]{Sellwood2010,Beuther2017}. Since we only consider emission from the end of the bar, the radial streaming velocity component must go to zero, and the resultant azimuthal velocity component, $v_{\phi}^B$, is
\begin{equation} 
v_{\phi}^B \sim \left( 1 - \frac{1 - q_{\phi}^2}{4 q_{\phi}^2} \right) V_C ,
\end{equation} 
where $q_{\phi}$ is the axial ratio of the bar potential. We use $1 - q_{\phi} \simeq \frac{1}{3}(1 - q)$ from \citet[][p.\,77]{Binney2008}, where $q = 0.3 - 0.4$ is the axial ratio of the density distribution for the Milky Way bar \citep{Bissantz2002}. We adopt $q = 0.35 \pm 0.05$ and find that $v_{\phi}^B = 0.84 \pm 0.02\,V_C$. Observed along the line of sight, this corresponds to a velocity of $57 \pm 4$\,\kms which is near our observed 45\,\kms velocity component.

\begin{figure}
  \centering \vspace{-10pt}
  \includegraphics[width=0.51\textwidth]{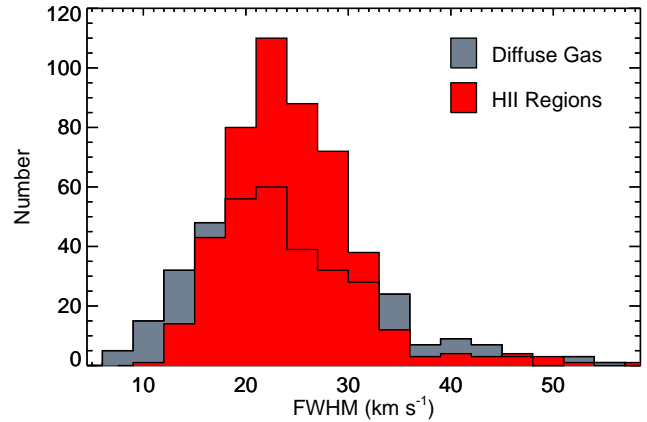}
  \caption{Distribution of FWHM line widths of directions coincident with discrete \hii\ regions and the DIG. There is no significant difference between the average line widths of \hii\ regions and the DIG (see text), indicating that turbulence does not play a major role in altering the observed velocity of the diffuse gas.\label{fig:hiivsdiff}}
\end{figure}

Although the above method describes the observed velocity components near $\ell \sim 30\degree$ fairly well, the assumption that the gas is located at the end of the Galactic bar breaks down when considering gas emission from the $\ell \sim 24\degree$ region further within the bar where we observed a similar velocity distribution. To describe the kinematics of the gas at this location, we must include radial streaming motions along the bar  \citep{Sellwood2010,Beuther2017} which can be estimated by
\begin{equation} 
v_{r}^B \sim \frac{2}{3} \left( 1 - \frac{1 - q_{\phi}^2}{4 q_{\phi}^2} \right) V_C .
\end{equation} 
We repeat the analysis above for the $\ell \sim 24\degree$ region, and find an unperturbed velocity component along the line of sight of 96\,\kms, and a perturbed velocity component of $7 \pm 2$\,\kms. In theory, shocks and turbulence could increase the latter to match our observed 45\,\kms emission. While we can not quantify the amount of turbulence in the DIG directly, we can compare the observed hydrogen recombination line widths at the diffuse directions with the line widths of directions coincident with discrete \hii\ regions. Assuming the same electron temperature, differences in line widths should trace the relative strength of turbulence between these directions. We find, however, no statistically significant difference of line widths between directions coincident with \hii\ regions (${\rm FWHM} = 24.5 \pm 6.4$\,\kms) and our diffuse directions (${\rm FWHM} = 23.7 \pm 8.8$\,\kms). We show the corresponding FWHM line width distributions in Figure~\ref{fig:hiivsdiff}. This suggests that turbulence does not play a significant role in altering the observed velocity of the gas. As a result, the large difference of the derived 7\,\kms velocity component to our observed 45\,\kms emission makes it doubtful whether interacting gas clouds at a single distance near $\ell \sim 24\degree$ could result in the observed velocity distribution.

\begin{figure*}
  \centering
  \includegraphics[width=1.05\textwidth]{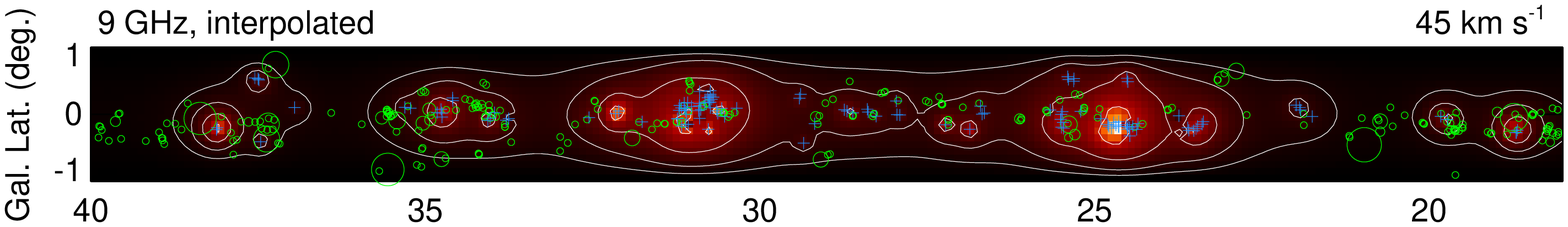}
  \includegraphics[width=1.05\textwidth]{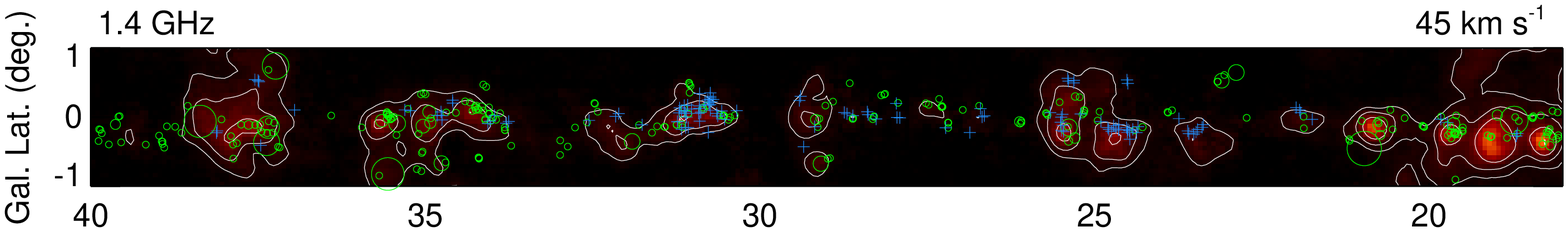}
  \includegraphics[width=1.05\textwidth]{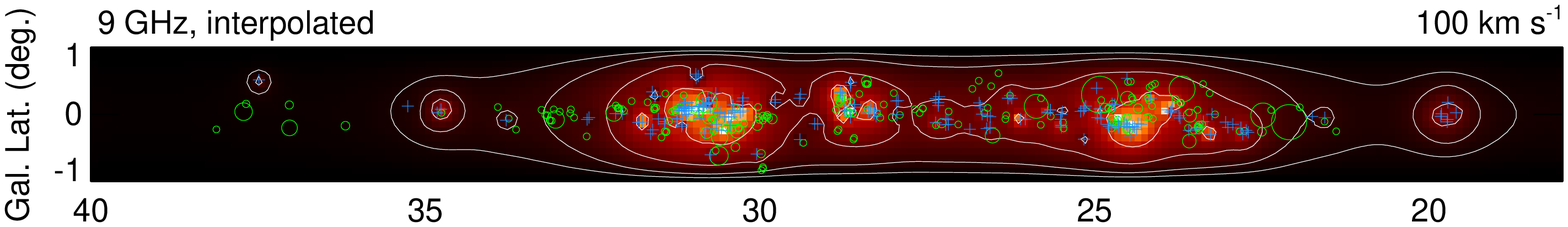}
  \includegraphics[width=1.05\textwidth]{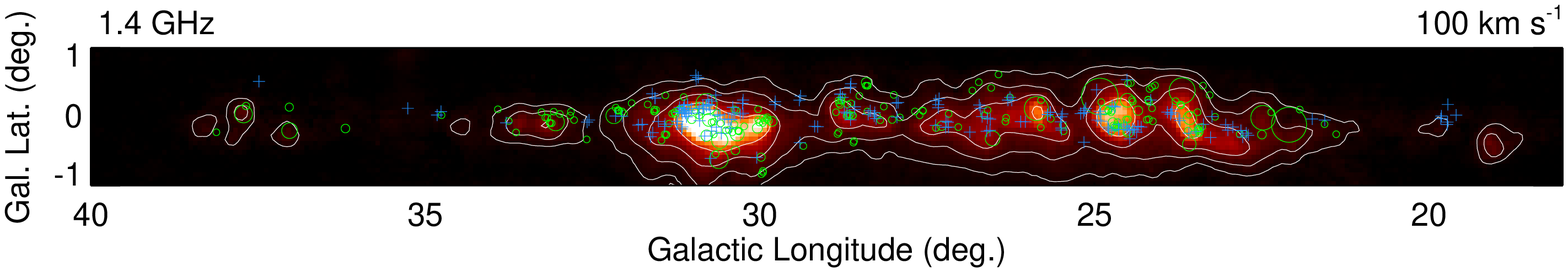}
  \caption{RRL emission over our observed longitude range. Top panel: Image of the diffuse gas emission for the $45$\,\kms\ velocity component. This image was made by interpolating our irregularly gridded 9\,GHz RRL data. The color scale shows the emission ranging linearly from 0 to 30\,mK, and the contours are at 2, 4, 8, and 16\,mK. The locations of the diffuse emission in this velocity range are marked by blue crosses, and green circles mark \hii\ regions from the velocity range indicated. The \hii\ region circle sizes and locations are from \citet{anderson14}, with a minimum circle size of 0.1\degree for better visibility. Second from top: Parkes 1.4\,GHz RRL survey map, integrated for the $45$\,\kms\ velocity component \citep{alves15}. Unlike our interpolated image above, the 1.4\,GHz data includes emission from discrete \hii\ regions. The color scale shows the emission ranging linearly from 0 to 200\,mK, and the contours are at 15, 30, 60, and 120\,mK. Third from top: Same as top panel, but for the $100$\,\kms\ velocity component. Bottom panel: Parkes RRL survey map for the $100$\,\kms\ velocity component (scale and contours equivalent to second panel from top).  \label{fig:diffuse} }
\end{figure*}

Although the simple model discussed above suggests that interacting gas clouds can not account for our observed data, a more thorough numerical analysis would be required to confirm this result. \citet{Renaud2013,Renaud2015} developed a hydrodynamical simulation of a Milky Way-like galaxy which includes star formation and stellar feedback through photoionization, radiative pressure and supernovae. They find that the leading edges of bars are favorable locations for converging gas flows and shocks. A similar model, focusing on bar kinematics in particular, may provide more insight towards the interaction processes near the bar-spiral arm interface.

\subsection{Intensity and Distribution of the DIG}
Our database of RRL parameters from the HRDS also allows us to investigate the spatial distribution of the DIG in the plane of the sky. Using our irregularly gridded data points, we examine the diffuse gas separately for the two velocities, $45$\,\kms\ and $100$\,\kms. We create maps of the DIG in these two velocity ranges by interpolating the irregularly-spaced grid of 233 points to create pixels $6\arcmin$ square.  We do this by first performing a Delauney triangulation (using the IDL program ``qhull'') and then create an \lb\ map of the RRL intensity from the DIG using inverse distance weighting (using the IDL program ``griddata''). This method has the advantage that the maximum and minimum values in the interpolated surface can only occur at sample points. We assume that the top and bottom edges of the map ($b = \pm 1\degree$) have zero intensity to ensure that the emission is constrained in latitude. We show these images in Figure~\ref{fig:diffuse} for the two velocity ranges. We also show in Figure~\ref{fig:diffuse} the 1.4\,GHz \hi\ Parkes All-sky survey RRL map \citep{alves15} averaged over the velocity ranges of the $45$\,\kms\ and $100$\,\kms components for comparison. The green circles in Figure~\ref{fig:diffuse} show the locations of discrete \hii\ regions cataloged by \citet[][Version 1.4]{anderson14} that are within the velocity range of interest, while the gray crosses show locations where the DIG was detected within the velocity range.

\begin{figure*}
  \centering
  \includegraphics[width=1.04\textwidth]{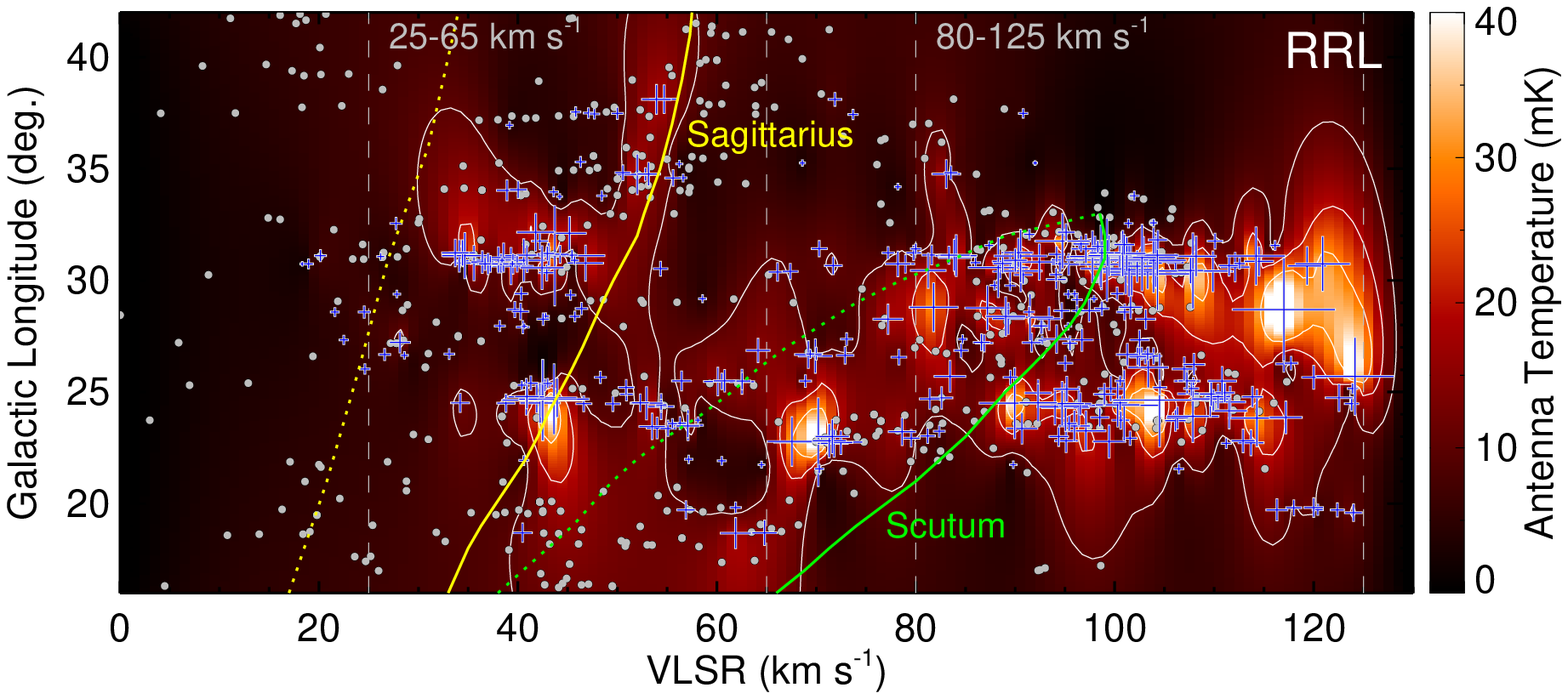}
  \includegraphics[width=1.04\textwidth]{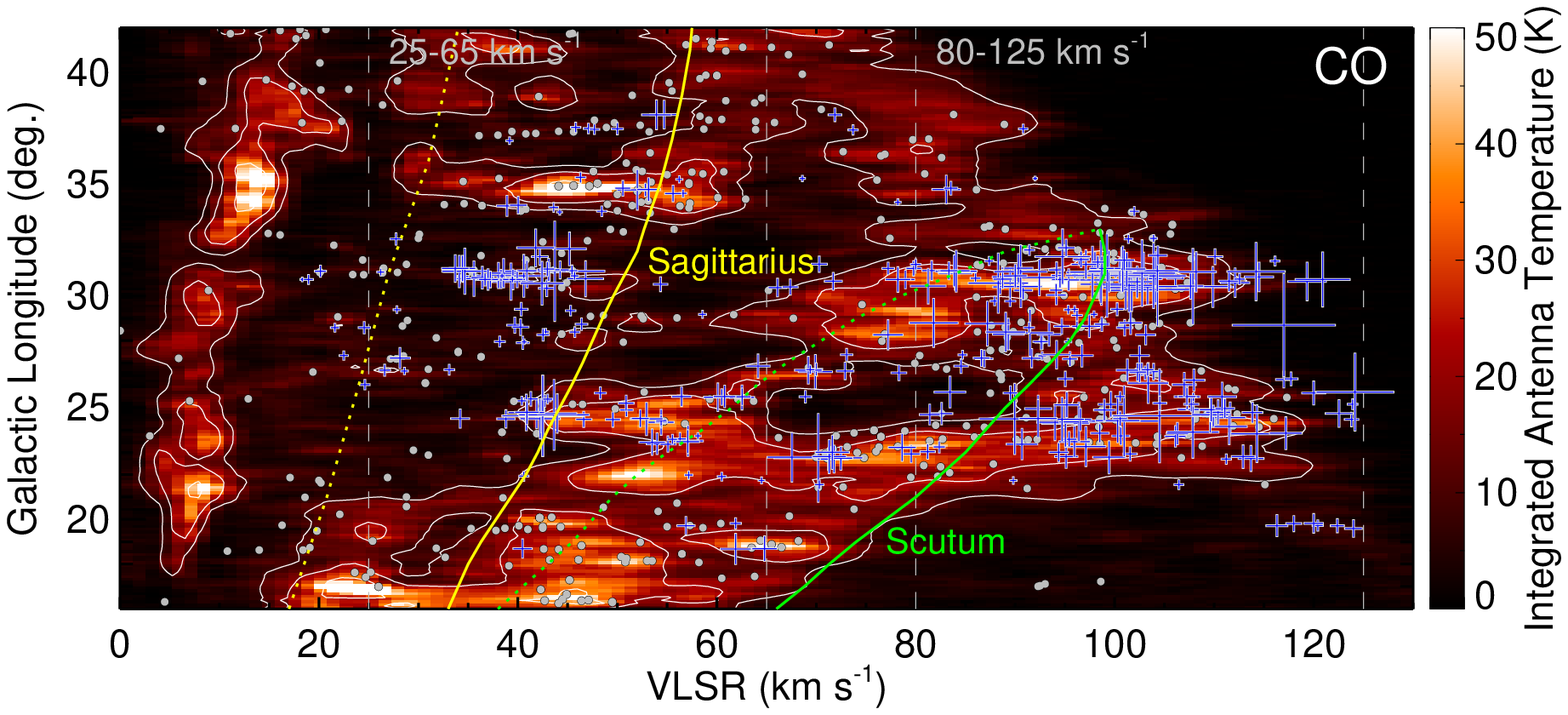}  
  \caption{Top: Longitude-velocity diagram of the DIG. The white contours are at 10, 20, and 30\,mK. The blue crosses denote the directions at which RRL spectra were taken and the cross sizes are proportional to the fitted H RRL intensities. The gray dots indicate discrete \hii\ regions. The Sagittarius and Scutum spiral arms are traced in yellow and green, respectively \citep[data from][]{Reid2016}. The dashed vertical lines indicate the two dominant velocity ranges: the 45\,\kms component at 25-65\,\kms and the 100\,\kms component at 80-125\,\kms. Bottom: Same, for Galactic $^{12}$CO emission, integrated in the range from $-$1\degree to $+$1\degree in latitude \citep[see][]{Dame1987}. Contours are at 10, 20, and 40\,K. \label{fig:lv}}
\end{figure*}

Using the same data set of RRL parameters, we explore the velocity distribution of the DIG in more detail. We create a longitude-velocity diagram of the DIG by interpolating between our grid points (Figure~\ref{fig:lv}, top panel), and assume that the velocity edges of the diagram (at 0\,\kms and 130\,\kms) have zero intensity so that the emission is constrained in velocity space. This assumption appears valid, since we did not detect any RRL components outside of this velocity range. In fact, our smallest and largest detected velocities at 18\,\kms and 124\,\kms, respectively, are well within this range. For comparison, we also show a longitude-velocity diagram of $^{12}$CO used to trace molecular clouds \citep[Figure~\ref{fig:lv}, bottom panel; data from][]{Dame1987}.

\subsubsection{The 45\,km\,s$^{-1}$ Gas Component}
The pixel-by-pixel correlation of RRL intensity at $45$\,\kms between our maps and the \citet{alves15} data is poor (see Figure~\ref{fig:alves_corr}, top panel). Our emission towards W43 near $\ell \sim 30 \degree$ and the $\ell \sim 24\degree$ region is disproportionately large in the $45$\,\kms map, whereas we do not see strong emission near the map edge at $\ell \sim 19\degree$. This may be due to interpolation errors between our sparse RRL pointings in this velocity and longitude range. Our low number of pointings may also be the cause of some of the more extended RRL emission between W43 and the $\ell \sim 24\degree$ complex that is less pronounced in the \citet{alves15} data. This makes it challenging to distinguish between interpolation errors and actual diffuse gas below the Parkes 1.4\,GHz RRL survey's sensitivity threshold for the undersampled regions in our maps. Additionally, the beam size of $\sim 14\arcmin$ in the \citet{alves15} maps is too large to avoid \hii\ regions at locations where their number density is high. Thus, most of their emission towards W43 and the $\ell \sim 24\degree$ region must be caused by discrete \hii\ regions rather than the DIG. Overall, the total integrated intensity of our maps is 46\% larger in the $45$\,\kms component compared with the \citet{alves15} data. This perhaps indicates that we are more sensitive to the diffuse gas.

\subsubsection{The 100\,km\,s$^{-1}$ Gas Component}
\begin{figure}
  \centering \vspace{-9pt}
  \includegraphics[width=0.52\textwidth]{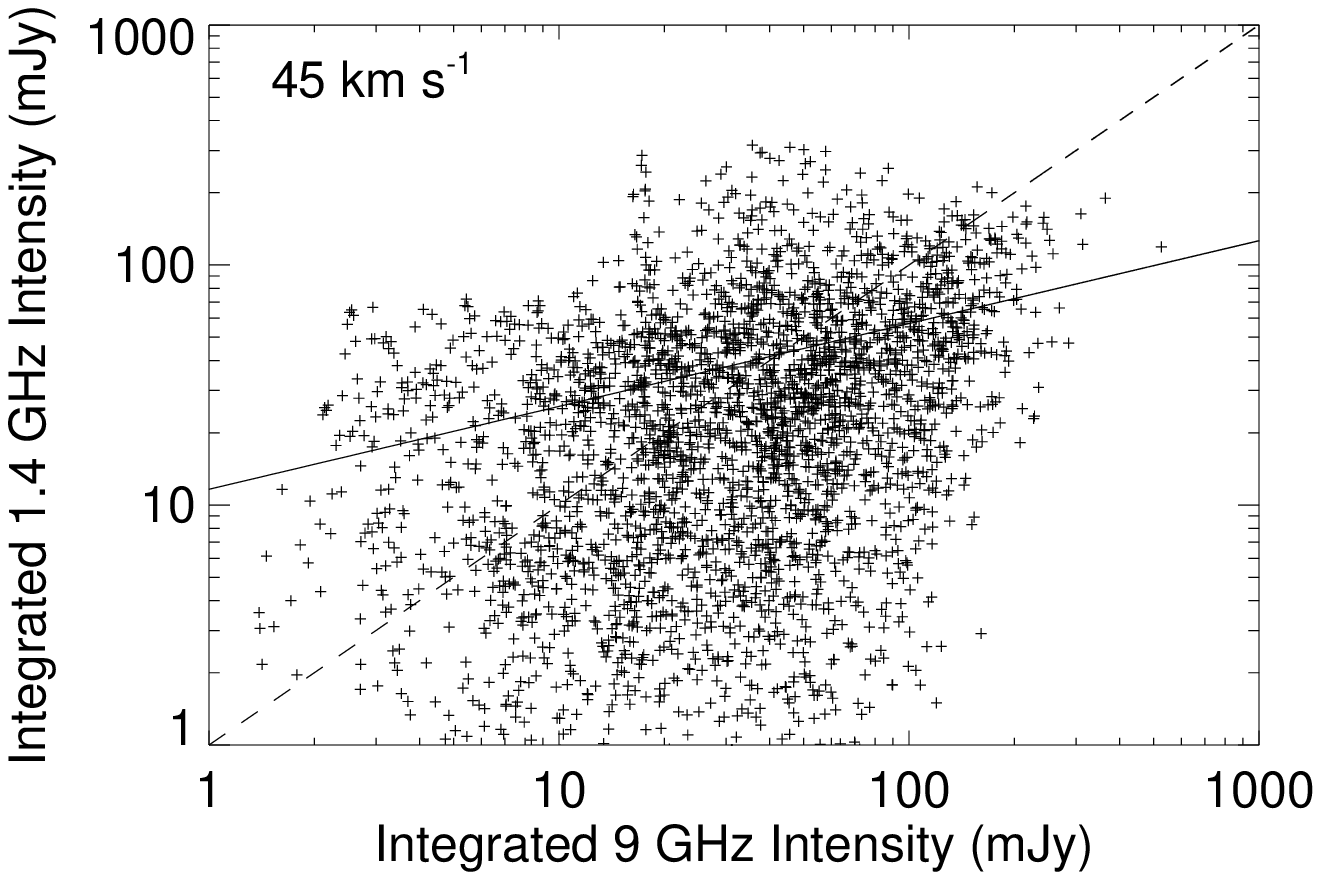}
  \includegraphics[width=0.52\textwidth]{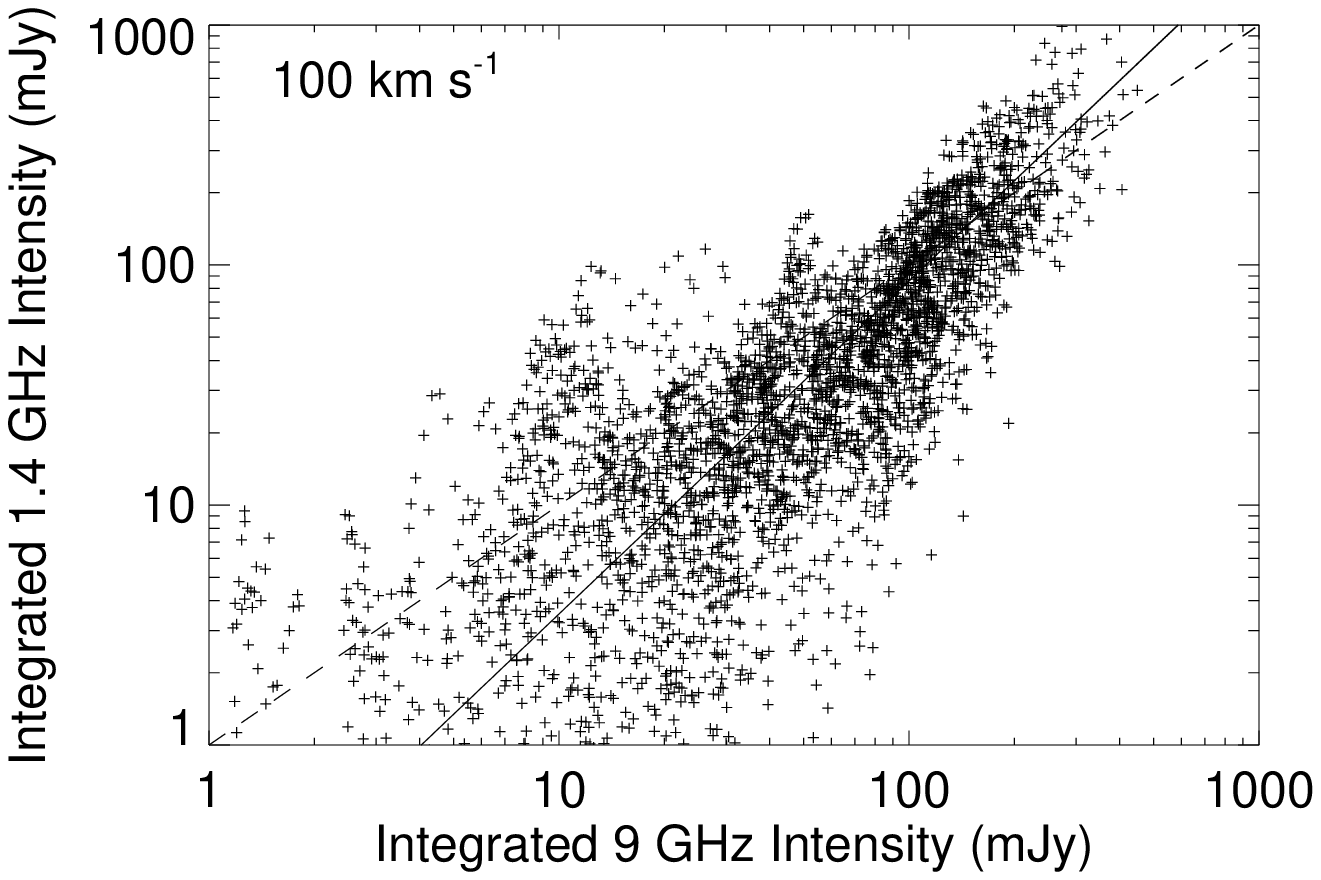}
  \caption{Top: Correlation of the diffuse gas emission between our interpolated data at 9\,GHz and the 1.4\,GHz map from \citet{alves15} for the $45$\,\kms\ velocity component. We assume a GBT gain of 2\,K\,Jy$^{-1}$ and integrate the intensity over a pixel size of 36\,sq.\,arcmin for both data sets. The solid line is a power-law fit of the form $y=ax^b$, with $a=11.6321 \pm 0.0295$ and $b=0.3452 \pm 0.0006$. The dashed line is a 1:1 relation. Bottom: Same, for the $100$\,\kms\ velocity component. The power-law fit parameters are $a=0.1440 \pm 0.0003$ and $b=1.3874 \pm 0.0004$. \label{fig:alves_corr}}
\end{figure}

While the interpolated $45$\,\kms map shows poor agreement with the \citet{alves15} data, our $100$\,\kms map is strongly correlated with the 1.4\,GHz RRL emission data (Figure~\ref{fig:alves_corr}, bottom panel). By-eye comparison of the two maps (Figure~\ref{fig:diffuse}) indicates that we are more sensitive to the diffuse gas component, especially at lower Galactic longitudes. The total integrated intensity of our data is 10\% larger in the $100$\,\kms component compared to \citet{alves15} (see Table~\ref{tab:par}). Figure~\ref{fig:lv} (top panel) shows that much of the $100$\,\kms emission from the DIG may be associated with the Scutum spiral arm. The higher velocities of the DIG compared to the Scutum arm may indicate that we are observing strong streaming motions in this direction \citep[see][]{Bania2012}. Alternatively, the DIG may be located near the inner edge of the Scutum tangency where it is falling into the arm's gravitational potential, as suggested by \citet{Langer2017}.

\begin{figure}
  \centering
  \includegraphics[width=0.48\textwidth]{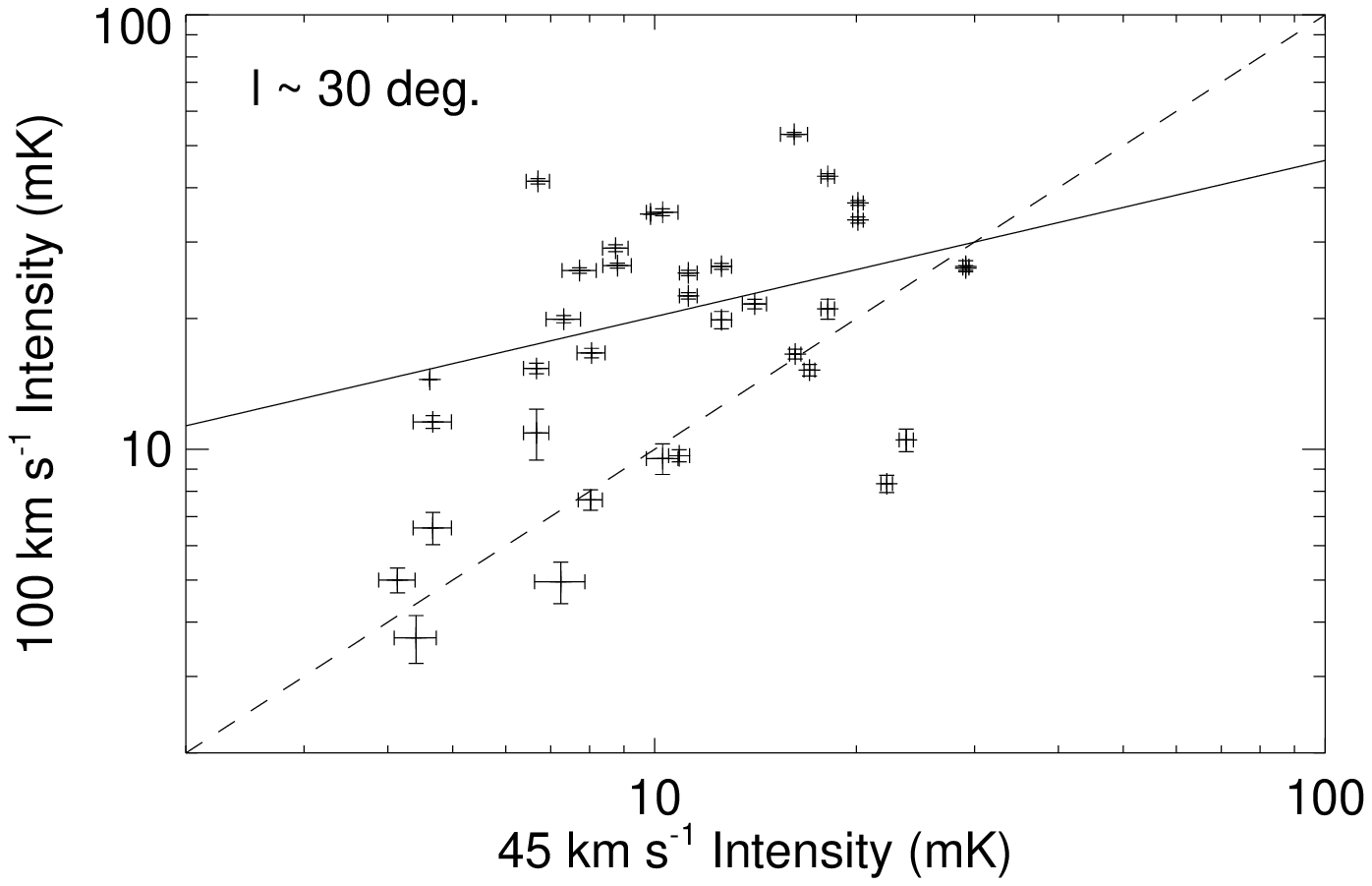}
  \includegraphics[width=0.48\textwidth]{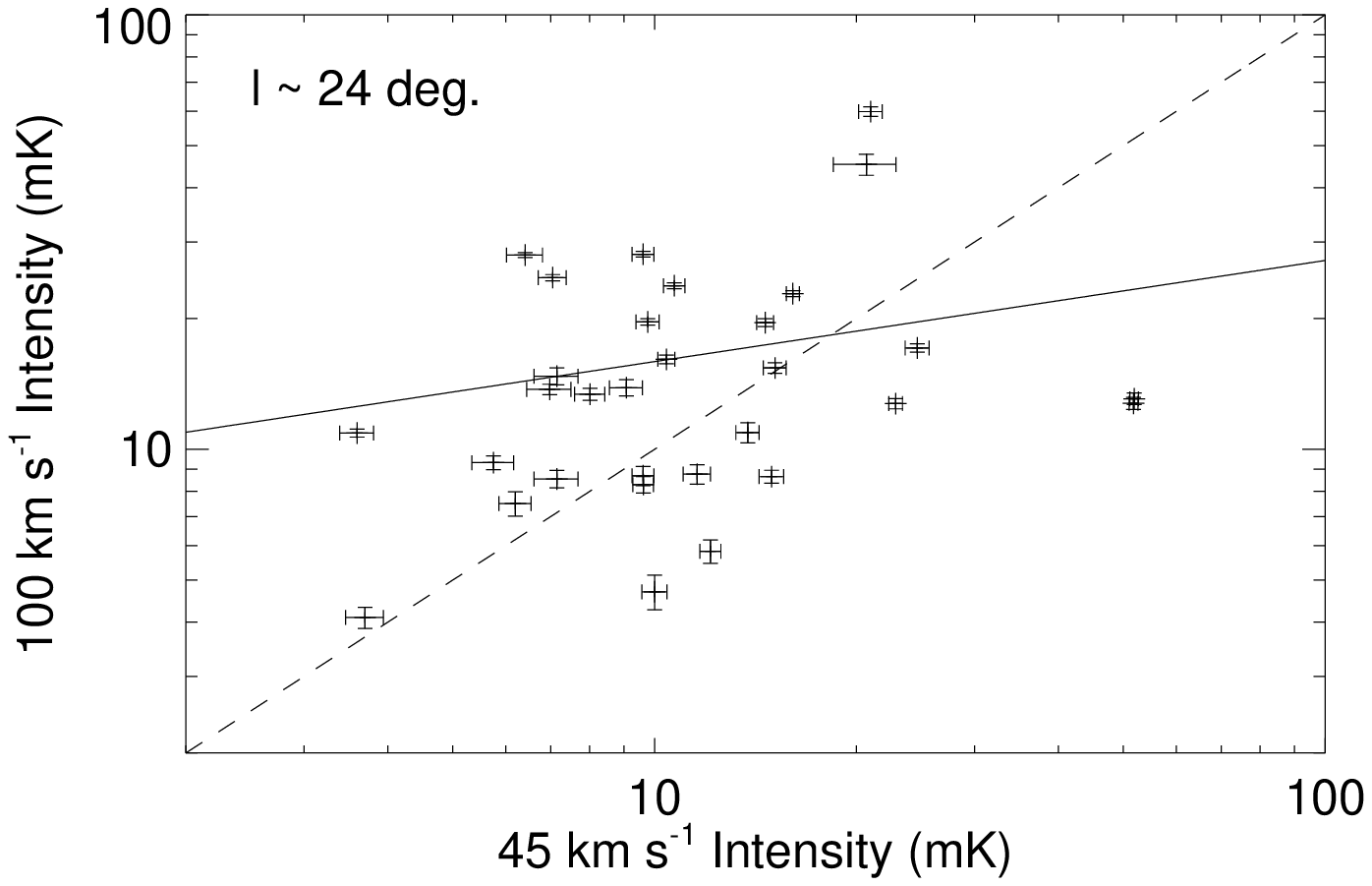}
  \caption{Top: Correlation between the observed $45$\,\kms\ and $100$\,\kms\ DIG intensities towards W43 ($\ell \sim 30\degree$). Each data point corresponds to a location where the RRL spectrum shows two hydrogen line profiles: one with a LSR velocity in the 25--65\,\kms range (the 45\,\kms component), and one with a LSR velocity in the 80--125\,\kms range (the 100\,\kms component). Error bars are $\pm$1$\sigma$. The solid line is a power-law fit of the form $y=ax^b$, with $a=8.8281 \pm 0.3453$ and $b=0.3595 \pm 0.0149$. The R$^2$ of the regression is 0.141. The dashed line is a 1:1 relation. Bottom: Same, for the $\ell \sim 24\degree$ region. Here, $a=9.3157 \pm 0.4008$ and $b=0.2327 \pm 0.0160$ (R$^2 = 0.058$).
 \label{fig:diffuse_correlation_w43}}
\end{figure}

The directions of strong emission in the two velocity ranges are slightly correlated, such that locations of strong emission from the DIG near $45$\,\kms\ mostly have strong emission near $100$\,\kms as well. The correlation is more significant towards the W43 region, whereas it is weak near $\ell \sim 24\degree$ as shown in Figure~\ref{fig:diffuse_correlation_w43}. This may suggest that the two velocity ranges towards W43 represent flows of interacting ionized gas (see \S 3.1.3), whereas the two velocity ranges towards the $\ell \sim 24\degree$ region could be caused by DIG emission at two distinct distances.

\subsection{H\,{\footnotesize I} and the Diffuse Gas}
It is uncertain whether a substantial amount of cold \hi\ gas can coexist with the diffuse ionized gas in regions with strong RRL emission from the DIG. If the radiation field in such regions is strong enough to ionize a large fraction of the gas, we may be able to observe a depletion in \hi\ at locations and velocities of strong DIG emission \citep{Miller1993, Domgoergen1994}.

This relationship has been probed by \citet{zurita02} who find that the \hi\ distribution does not correlate (or anti-correlate) at all with the diffuse H$\alpha$ emission tracing the ionized gas in the face-on galaxy NGC\,157. The angular resolution of their \hi\ map, however, does not match the better resolution of their H$\alpha$ map. As a result, they would not be able to resolve \hi\ depletion cavities much smaller than 1\,kpc. A previous study by \citet{Reynolds1995a} analyzed the same correlation for H$\alpha$-emitting \hi\ clouds in the Milky Way. They find that the neutral and ionized components in these clouds are likely spatially separated. Since they only observed a relatively small region of the sky away from the Galactic plane, their available sample size is limited.

\begin{figure*}
  \centering
  \begin{tabular}{cc}
  \includegraphics[width=0.48\textwidth]{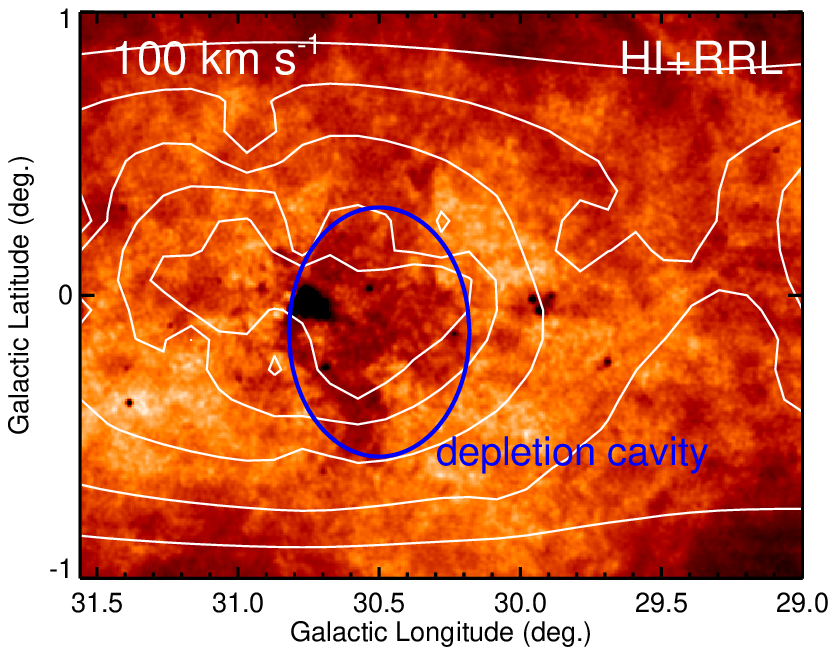} &
  \includegraphics[width=0.48\textwidth]{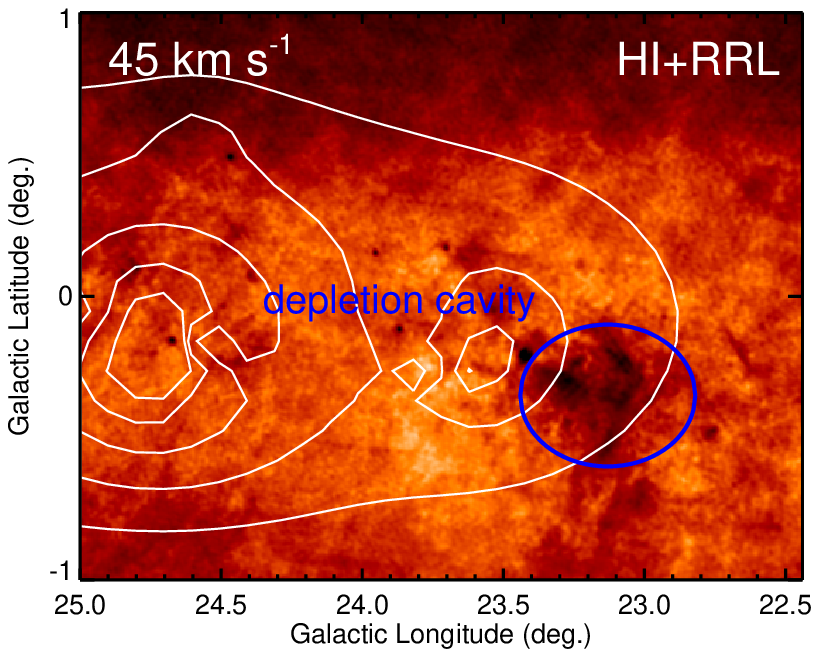} \vspace{-10pt} \\
  \includegraphics[width=0.48\textwidth]{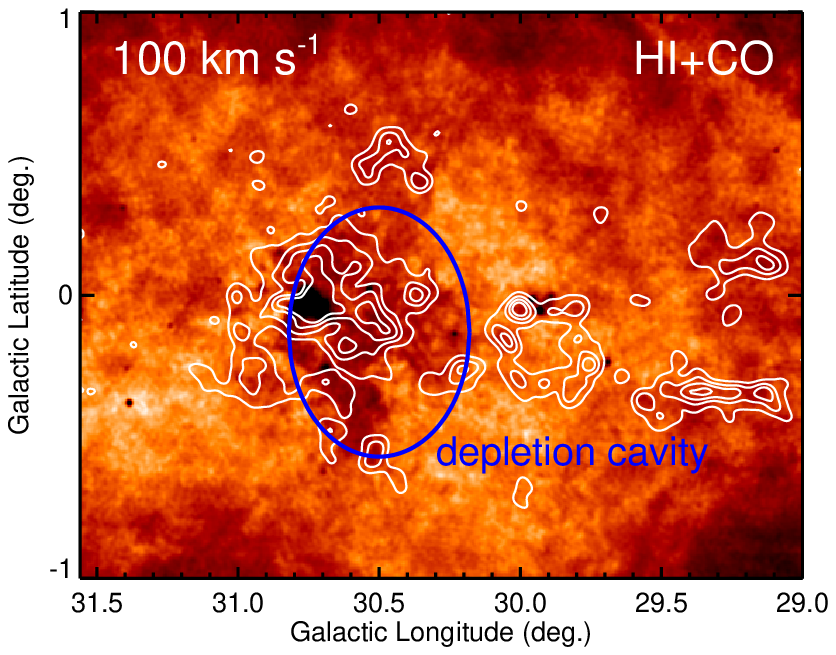} &
  \includegraphics[width=0.48\textwidth]{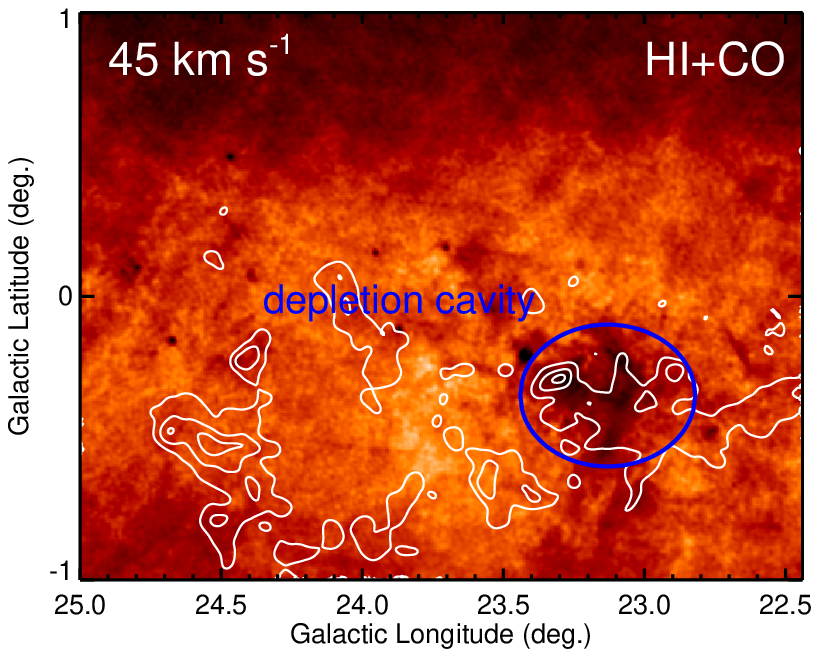} \\
  \end{tabular}
  \caption{Comparison of RRL emission from the DIG with \hi\ and CO emission reveals that deficiencies in \hi\ emission are more likely due to self-absorption rather than gas displacement by the DIG. Top left: The VGPS \hi map at 92\,\kms for the $29\degree < \ell < 31.5\degree$ region, linearly scaled from 0 to 150\,K brightness temperature. The white contours show our interpolated diffuse gas map at the $100$\,\kms\ velocity component at 4, 8, 12, 16, and 20\,mK. A deficiency in \hi emission is clearly visible at $\ell \sim 30.5\degree$, the location of W43. Top right: VGPS \hi map at 60\,\kms for the $22.5\degree < \ell < 25\degree$ region, with contours of our $45$\,\kms\ diffuse gas map and same increments as above. An \hi\ deficiency is visible at $\ell \sim 23\degree$ which is, however, not spatially correlated with the diffuse gas emission. Bottom left: Same as top left, but with blue GRS $^{13}$CO contours at the same velocity. Contours are at $0.5$, $1$, $1.5$, and $2$\,K antenna temperature. The CO emission is spatially correlated with the \hi\ bubble above, indicating that the deficiency in \hi\ emission may be caused by \hi\ self-absorption. Bottom right: Same, for the $22.5\degree < \ell < 25\degree$ region at 60\,\kms. \label{fig:hi_bubble}}
\end{figure*}

We use the VGPS \hi\ data to test whether regions with strong  emission from the DIG show a deficiency in \hi. The VGPS data cubes have a spatial resolution of $1\arcmin \times 1\arcmin$ and a spectral resolution of 1.56\,\kms. For the W43 region near $\ell \sim 31\degree$ we find an apparent \hi depletion cavity at 92\,\kms (Figure~\ref{fig:hi_bubble}, top left panel) which is consistent with our strong DIG emission shown in Figure~\ref{fig:diffuse}. We find a similar depletion cavity for the $\ell \sim 23\degree$ region at 60\,\kms (see Figure~\ref{fig:hi_bubble}, top right panel) which is, however, offset by $\sim$$1.5\degree$ from the strong DIG emission seen in the $45$\,\kms\ component near $\ell \sim 24.5\degree$ (Figure~\ref{fig:diffuse}). Our large number of RRL pointings near the $\ell \sim 24\degree$ region makes it unlikely that this offset is an artifact from our interpolation algorithm. We did not find any other strong \hi\ depletion cavities in the velocity ranges of significant emission from the DIG for the sky zones in Figure~\ref{fig:hi_bubble}.

Comparison of our RRL data with $^{13}$CO maps casts doubt on whether strong ionized gas emission is usually spatially associated with a deficiency in \hi. Using $^{13}$CO GRS data, we can determine for both our regions whether these deficiencies are caused by \hi\ self-absorption or due to an actual lack of \hi\ gas. \hi\ self-absorption, first described in detail by \citet{knapp74}, is usually correlated with CO emission features \citep{burton78, Garwood1989}. We find substantial $^{13}$CO emission at the two directions and velocities (Figure~\ref{fig:hi_bubble}, bottom panels). In fact, the integrated $^{13}$CO emission found at $\ell \sim 31\degree$ and $\ell \sim 23\degree$ is among the strongest within the range of the GRS. This suggests that the lack in \hi\ emission is caused by \hi\ self-absorption and is not due to an actual deficiency in \hi\ gas.

\begin{figure*}
  \centering
  \includegraphics[width=1\textwidth]{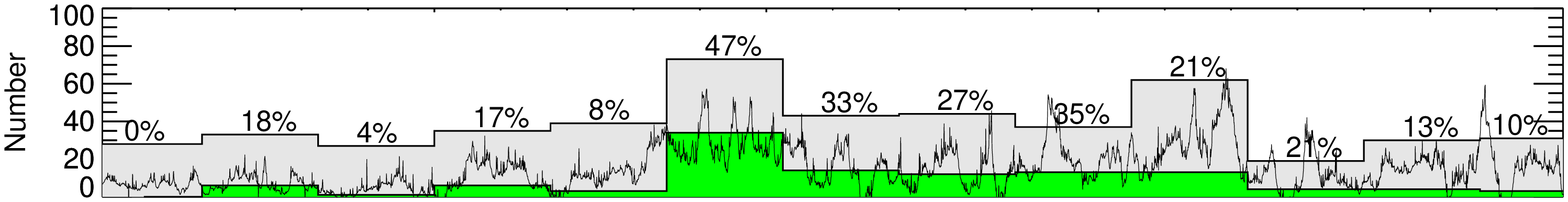}
  \includegraphics[width=1\textwidth]{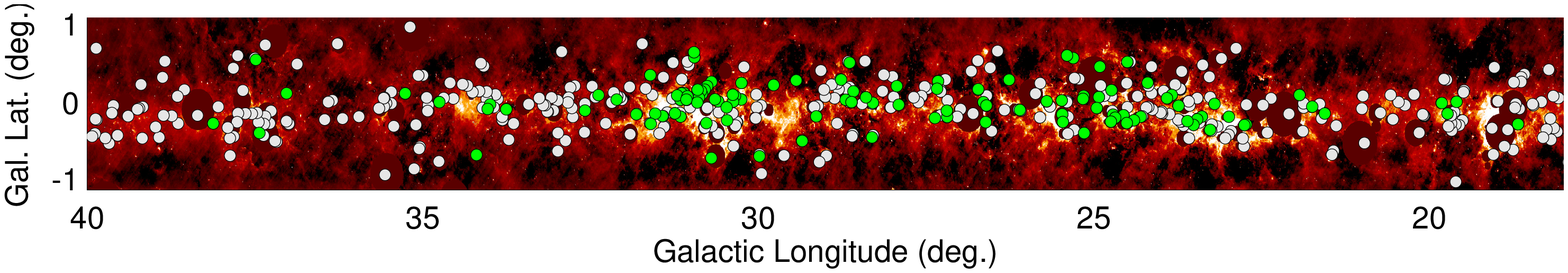}
  \caption{Locations of multiple-velocity \hii\ regions. In the bottom panel, the background image is the point-source subtracted 8.0\,\micron\ {\it Spitzer} GLIMPSE emission, on a square-root scale ranging from -5 to 50\,MJy\,sr$^{-1}$. Multiple-velocity \hii\ regions are shown in green and single velocity \hii\ regions are shown in gray. Emission around \hii\ regions is blanked out according to their positions and sizes from the WISE catalog \citep{anderson14}. In the top panel we show histograms of the number of single-velocity \hii\ regions (gray) and multiple-velocity \hii\ regions (green). The numbers above the histogram indicate the percentage of multiple-velocity regions over the given longitude range. The black curve in the top panel shows the 8.0\,\micron\ GLIMPSE emission from the bottom panel integrated over all latitudes. There is a correlation between the number of \hii\ regions and the 8.0\,\micron\ intensity (see Figure~\ref{fig:multvel_hist_corr}). \label{fig:multvel_pos_glimpse}}
\end{figure*}

\subsection{Correlation with 8.0\,$\mu$m Intensity}
Emission from polycyclic aromatic hydrocarbons (PAHs) within the 8.0\,\micron\ band is usually caused by softer ultra-violet (UV) radiation than that responsible for RRL emission from \hii\ regions. For example, \citet{Robitaille2012} show, using radiative transfer models, that most of PAH heating is provided by B stars, compared to RRL emission typically caused by O stars. While the 8.0\,\micron\ emission is often associated with strong PDRs surrounding discrete \hii\ regions, there exists significant PAH emission that originates from the diffuse gas without nearby \hii\ regions. Below, we analyze this ``diffuse" PAH emission and its relation to the DIG. In the bottom panel of Figure~\ref{fig:multvel_pos_glimpse} we show a map of the point-source subtracted 8.0\,\micron\ {\it Spitzer} GLIMPSE emission \citep{Benjamin2003, Churchwell2009}. Since there is also strong 8.0\,\micron\ emission from discrete \hii\ regions, we blank out these regions based on their corresponding positions and sizes from the WISE catalog, Version 1.4 \citep{anderson14}. In the upper panel of Figure~\ref{fig:multvel_pos_glimpse} we show histograms of the location of single-velocity and multiple-velocity \hii\ regions. 

\begin{figure}
  \centering \vspace{-8pt}
  \includegraphics[width=0.49\textwidth]{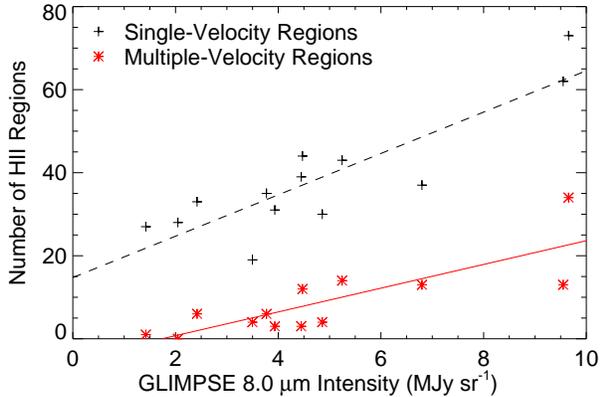}
  \caption{Correlation between 8.0\,\micron\ emission from PAHs and the number of \hii\ regions. Each data point represents one histogram bin from Figure~\ref{fig:multvel_pos_glimpse}, averaged over the entire $|b|<1\degree$ latitude range. The dashed line is a linear fit of the form $y=a+bx$ for the single-velocity regions, with $a=14.743 \pm 0.6056$ and $b=4.9803 \pm 0.1127$. The solid line is the fit for the multiple-velocity regions, with $a=-4.9657 \pm 0.6057$ and $b=2.8588 \pm 0.1127$. The correlation indicates that a significant amount of soft UV radiation is present near discrete \hii\ regions. \label{fig:multvel_hist_corr}}
\end{figure}

We observe a correlation between the location of discrete \hii\ regions and the intensity of the 8.0\,\micron\ emission (see Figure~\ref{fig:multvel_hist_corr}). Both of these diffuse emission components should be caused by UV photons leaking from the discrete \hii\ regions. The PAHs responsible for the 8.0\,\micron\ emission are destroyed in the hard UV radiation within an \hii\ region \citep{voit92, povich07}, but can survive where the radiation field is softer, i.e.~in the diffuse ISM. The correlation therefore suggests that either a significant amount of the (soft) UV photons responsible for the 8\,\micron\ emission is leaking from the discrete \hii\ regions or that the harder UV radiation produced by the O stars softens as it escapes into the ISM. Such a radiation softening has been suggested by \citet{Reynolds1995} and was recently observed indirectly for the compact \hii\ region NGC\,7538 \citep{Luisi2016}.

We also observe a correlation between the hard UV radiation field within the DIG and the softer UV radiation field responsible for PAH emission. Our method probes the radiation field strengths by using the observed RRL emission from the DIG and the 8.0\,\micron\ emission intensities as a proxy for the diffuse hard UV and soft UV radiation field strengths, respectively. A correlation between the intensities of these two emission components indirectly tests for a correlation between the radiation fields. We determine the diffuse 8.0\,\micron\ background by integrating the 8.0\,\micron\ flux in a circular $41\arcsec$ aperture centered at each of the 135 off-target directions. By only using the off-target directions, we ensure that we are only sampling the diffuse 8.0\,\micron\ background, and not the emission associated with \hii\ region PDRs. We then compute the fluxes using our Kang software\footnote{http://www.bu.edu/iar/files/script-files/research/kang/}. Kang is an astronomical visualization and analysis package written in IDL.  Its relevant functionality here is that it can compute aperture photometry measurements using arbitrary aperture shapes.

\begin{figure}
  \centering \vspace{-16pt}
  \includegraphics[width=0.48\textwidth]{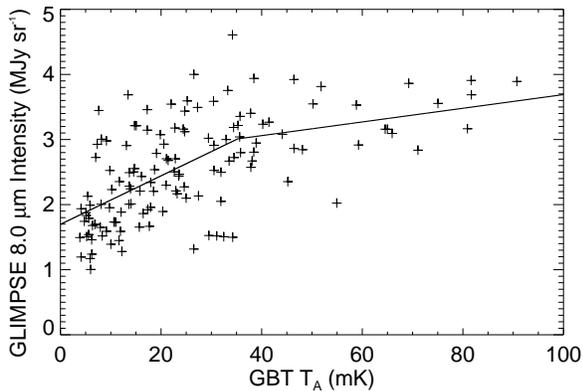}
  \caption{Correlation between 8.0\,\micron\ emission from PAHs and RRL emission from the DIG. The RRL emission is from our pointed GBT observations of the diffuse gas. We average the point-source subtracted 8.0\,\micron\ \emph{Spitzer} GLIMPSE intensity over a $41\arcsec$ radius aperture (the beam size of the RRL observations) at the directions of our GBT observations. The two straight lines show linear fits, and indicate that the intensity of the far-UV radiation responsible for the RRL emission is correlated with the softer UV radiation responsible for the 8.0\,\micron\ emission. The break between the fits is at GBT $T_A = 35$\,mK which suggests that above this temperature PAHs are destroyed in highly energetic radiation fields, leading to a saturation in IR emission. \label{fig:glimpse_diffuse}}
\end{figure}

We compare the diffuse 8.0\,\micron\ emission with the integrated intensity from all hydrogen RRLs at all velocities detected at an off-target direction. We plot in Figure~\ref{fig:glimpse_diffuse} the correlation between the diffuse 8.0\,\micron\ emission and the integrated RRL intensity. Figure~\ref{fig:glimpse_diffuse} indicates that the hard UV radiation is correlated with the softer UV radiation, as one would expect, although there is quite a large scatter.

Furthermore, the correlation appears to change above 35\,mK. This may be due to the destruction of the PAH molecules that are largely responsible for the diffuse 8.0\,\micron\ emission. A similar effect has been observed by \citet{Lebouteiller2011} in the Galactic \hii\ region NGC\,3603. Direct PAH destruction typically requires photons with energies $>$20\,eV which exist in sufficient numbers only in the most energetic radiation fields (e.g., within \hii\ regions). The binding energy of H atoms to PAHs, however, is only $\sim$4.8\,eV \citep{voit92}. Thus, even less energetic radiation fields can contribute to PAH dissociation. The relatively strong hydrogen ionizing radiation field ($\geq 13.6$\,eV) within regions with substantial RRL emission from the DIG must therefore partly be responsible for PAH dissociation. Considering that this effect limits the abundance of PAHs in these regions, it comes as no surprise that we observe a saturation in PAH emission.
 
Clearly, further study of this correlation is required. Most helpful would be additional pointings toward cleaner sight lines where the source of the UV photons can be more easily be determined. A more direct measurement of the radiation field strength is difficult. Recently, \citet{Stock2017} described a technique to estimate the UV field intensity using the ratio between two PAH spectral components at 7.6\,\micron\ and 7.8\,\micron. While this technique has not yet been applied to the diffuse ISM, it may prove useful in constraining properties of the radiation field outside \hii\ regions.

\section{Summary}
Here, we analyze the DIG using hydrogen RRL emission line spectra in the range $\ell=18\degree$ to $40\degree$ which are either devoid of \hii\ region emission or have multiple velocity components. Our data set is comprised of 353 RRL emission line components from the DIG. These allow us to determine the intensity and distribution of the diffuse gas. 

We find that the DIG is spatially concentrated in two areas near $\ell=31\degree$ and $\ell=24\degree$, with two dominant velocity components ($45\,\kms$ and $100\,\kms$) in each of the areas. We investigate the KDA for the two velocity ranges and conclude that the $100\,\kms$ component has a Galactocentric distance of $\sim$6\,kpc, corresponding to the location of W43. This suggests that much of the $100\,\kms$ gas is associated with W43. The origin of the $45$\,\kms\ component is less clear. The intensity of the emission in the two velocity ranges is slightly correlated near $\ell=31\degree$, which may imply that both velocity components originate at a single distance. In this case, the $45$\,\kms\ component may arise from complex streaming motions near the end of the Galactic bar. For the $\ell=24\degree$ region, however, it is unlikely that the observed velocity components are due to this effect. As an alternative, we suggest that the $45$\,\kms\ emission may have its origin at a Galactocentric distance of $\sim$12\,kpc, or a combination of both. Unfortunately, our current data are insufficient to clearly distinguish between these cases, a problem which may be investigated in future work. Future work may also explore in more detail the connection between the DIG observed in RRL emission, the more diffuse component observed in H$\alpha$, and the different environments these data are tracing.

Since regions with strong RRL emission from the DIG may show a deficiency in \hi, we examine data from the VGPS for \hi\ depletion cavities. We find such a bubble in \hi\ emission for the W43 region at $\ell \sim 31\degree$ at 92\,\kms. We also find a second bubble at $\ell \sim 23\degree$ and 60\,\kms which is, however, offset by $\sim$$1.5\degree$ from the direction of strong DIG emission at $\ell \sim 24.5\degree$. There is strong $^{13}$CO emission associated with these locations, suggesting that the deficiency in \hi\ emission is rather caused by \hi\ self-absorption than an actual lack of \hi\ gas.

The intensity of the RRL emission from the DIG is also correlated with the intensity of diffuse {\it Spitzer} GLIMPSE 8.0\,\micron\ emission, implying that the soft UV photons responsible for creating the infrared emission have a similar origin as the harder UV photons required for the RRL emission. The diffuse 8.0\,\micron\ emission appears to saturate at locations with the strongest RRL emission suggesting that the PAHs responsible for the 8.0\,\micron\ emission are destroyed by the radiation field in these regions.

\vspace{20pt} \acknowledgements
We thank Robert A.~Benjamin and Brian L.~Babler for providing us with the point-source subtracted 8.0\,$\mu$m data. Support for TVW was provided by the NSF through the Grote Reber Fellowship Program administered by Associated Universities, Inc./National Radio Astronomy Observatory. \nraoblurb 

\facility{Green Bank Telescope.}
\software{TMBIDL \citep{Bania2014}, Kang.}

\bibliographystyle{aasjournal} 
%\bibliography{all}
\bibliography{HII,Loren}

\end{document}